\documentclass[aps,prd,twocolumn,superscriptaddress,longbibliography,nofootinbib]{revtex4-2}
\usepackage{graphicx,longtable,mathrsfs,color,array}
\usepackage[hidelinks]{hyperref}
\usepackage{amsmath}
\usepackage[dvipsnames]{xcolor}
\usepackage{mathrsfs}
\usepackage{amssymb}
\usepackage{orcidlink}
\usepackage{comment}
\usepackage[normalem]{ulem}

\def\prl{Phys. Rev. Lett.}
\def\prd{Phys. Rev. D}

\def\apj{Astrophys. J.}

\def\aap{Astronomy and Astrophysics}
\def\aaps{Astronomy and Astrophysics Suppl.}

\def\pau_p{Prog. Theor. Phys.}

\def\mnras{Mon. Not. R. Astron. Soc.}

\def\apss{Astrophys. Space Sci.}
\def\physrep{Phys. Rep.}

\def\sovast{Soviet Ast.}

\begin{document}

\title{Accretion onto a moving black hole for stiff equations of state}

\author{Elliott Ewell}
\affiliation{Department of Physics and Astronomy, Bowdoin College, Brunswick, Maine 04011, USA}

\author{Peter K. Harris}
\affiliation{Department of Physics and Astronomy, Bowdoin College, Brunswick, Maine 04011, USA}

\author{Thomas W.~Baumgarte\orcidlink{0000-0002-6316-602X}}
\email{tbaumgar@bowdoin.edu}
\affiliation{Department of Physics and Astronomy, Bowdoin College, Brunswick, Maine 04011, USA}

\author{Stuart L.~Shapiro\orcidlink{0000-0002-3263-7386}}
\email{slshapir@illinois.edu}
\affiliation{Department of Physics, University of Illinois at Urbana-Champaign, Urbana, Illinois 61801}
\affiliation{Department of Astronomy and NCSA, University of Illinois at Urbana-Champaign, Urbana, Illinois 61801}

\begin{abstract}
We revisit accretion of an asymptotically uniform fluid onto a moving Schwarzschild black hole, the relativistic analog of Bondi-Hoyle-Lyttleton accretion.  Expanding on previous treatments, we focus on relativistic accretion of stiff fluids with adiabatic indices $\Gamma \geq 5/3$, which is particularly relevant for the treatment of primordial black holes captured by neutron stars.  We perform numerical simulations to systematically explore the dependence of the 
steady-state accretion and drag rates on the asymptotic sound speed $a_\infty$ and relative speed $v_\infty$, and identify different regimes, both subsonic and supersonic, in which these rates either increase or decrease with respect to the spherical accretion rate for $v_\infty = 0$.  We propose simple analytical approximations that capture the qualitative features observed in the dependence of the accretion and drag rates on $a_\infty$ and $v_\infty$.
\end{abstract}

\maketitle
%
\section{Introduction and Motivation}
\label{sec:intro}
%


Accretion processes are ubiquitous in astrophysics, and play a particularly important role in the physics of compact objects (see, e.g., \cite{ShaT83,FraKR02} for textbook treatments). Accordingly, numerous authors have studied accretion onto black holes and neutron stars in both Newtonian and fully relativistic gravity and for both subsonic and supersonic flow. Studies have explored many different effects on the accretion rate, including those of magnetic fields, black-hole rotation, morphology of the accretion flow, and radiation.

Expanding on the treatment by several previous authors (e.g.~\cite{ShiMTS85,PetSST89,FonI98a,FonI98b}; see also Table I in \cite{FogGR05} for a summary), we revisit in this paper a more limited problem, namely that of adiabatic accretion onto a Schwarzschild black hole moving in an ideal gas that is asymptotically homogeneous and at rest. (see also \cite{Edg04} for a review and \cite{ShaT83,RezZ13} for textbook treatments).  Specifically, we solve the equations of relativistic hydrodynamics to explore the dependence of the accretion and drag rates on the relative (asymptotic) speed $v_\infty$ of the black hole for representative soft and stiff equations of state (EOSs).

In an early analytical treatment of accretion onto black holes, Hoyle and Lyttleton \cite{HoyL39} considered the Newtonian accretion of dust with mass-density $\rho_{0\infty}$ onto a point-mass $M$ moving through the dust with speed $v_\infty$.  Throughout this paper, the subscript $\infty$ refers to asymptotic quantities measured at large distance from the black hole hole. The accretion rate $\dot M_{\rm HL}$ can then be computed from
\begin{equation} \label{eq:m_dot_HL_flux}
\dot M_{\rm HL} = \sigma v_\infty \rho_{0\infty}, 
\end{equation}
where $\sigma = \pi \zeta_{\rm crit}^2$ is the effective accretion cross-section and $\zeta_{\rm crit}$ the critical impact parameter separating those dust particles that end up getting accreted from those that do not.  Using simple Newtonian arguments, Hoyle and Lyttleton found
\begin{equation} \label{eq:zeta_crit_HL}
	\zeta_{\rm  crit} = \frac{2 G M}{v_\infty^2},
\end{equation}
which results in the Hoyle-Lyttleton accretion rate
\begin{equation} \label{eq:m_dot_HL}
	\dot M_{\rm HL} = \frac{4 \pi G^2 M^2 \rho_{0\infty}}{v_\infty^3}
\end{equation}
(see also \cite{BonH44} as well as \cite{Edg04} for a review).

Bondi \cite{Bon52} considered Newtonian, spherical, adiabatic accretion of gas onto a point-mass at rest with respect to the gas at large distances, and found
\begin{equation} \label{eq:m_dot_Bondi}
	\dot M_{\rm sph} = \frac{4 \pi \lambda G^2 M^2 \rho_{0\infty}}{a_\infty^3},
\end{equation}
where $a_\infty$ is the asymptotic sound speed and $\lambda$ an accretion eigenvalue typically of order unity.  For polytropic equations of state
\begin{equation} \label{eq:polytrope}
P = K \rho_0^{\Gamma},
\end{equation}
where $1 \leq  \Gamma \equiv 1 +1/n \leq 5/3$ is the adiabatic exponent, $n$ the polytropic index, and $K$ a constant,  these eigenvalues depend on $\Gamma$ only and are given by 
\begin{equation} \label{eq:lambda}
	\lambda = \frac{1}{4} \left( \frac{2}{5 - 3 \Gamma} \right)^{(5 - 3 \Gamma)/(2 \Gamma - 2)}~~(\mbox{Newt.},~1 < \Gamma \leq 5/3).
\end{equation}
A later relativistic treatment by Michel (\cite{Mic72}; see also \cite{ShaT83} for a textbook treatment that proves that the flow must be transonic) shows that fluid accretion onto a Schwarzschild black hole of mass $M$ is given by the same expression \eqref{eq:m_dot_Bondi}, except that the eigenvalues $\lambda$ now depend on $a_{\infty}$ in addition to $\Gamma$, that the rest-mass density $\rho_0$ should now be distinguished from the total mass-energy density $\rho$, and that the value of $\dot M$ should be interpreted as the rate of rest-mass accretion onto the black hole, as measured by an observer at rest at infinity, rather than the rate of increase in the gravitational mass. For spherical, steady state fluid accretion, the black hole's gravitational mass $M$ increases at a rate of
\begin{equation} \label{eq:M_grav_dot}
\dot M = h_\infty \dot M_0,
\end{equation}
where $h_\infty$ is the fluid's asymptotic enthalpy (see \cite{PetSST89,AguST21} as well as Appendix A in \cite{BauS24d}).  In (\ref{eq:M_grav_dot}) and in the following we denote the rest-mass accretion rate with $\dot M_0$ in order to distinguish it from the total mass-energy accretion rate $\dot M$.\footnote{Note, however, that $\dot M_0$ should not be interpreted as the time derivative of the black hole's rest mass, which is not defined.}

With the exception of one extreme-relativistic case discussed below, the exact rate of rest-mass accretion onto a black hole that moves through a gas with a nonvanishing asymptotic speed $v_\infty$ has not been found analytically.  In the absence of such an analytical expression, however, the similar functional form of \eqref{eq:m_dot_HL} and \eqref{eq:m_dot_Bondi} motivates an approximation 
\begin{equation} \label{eq:m_dot_can}
\dot M_{\rm can} = \frac{4 \pi \lambda G^2 M^2 \rho_{0\infty}}{(v_\infty^2 +  a_\infty^2)^{3/2}}
= M_{\rm sph} \left( \frac{ a_\infty^2}{v_\infty^2 + a_\infty^2} \right)^{3/2},
\end{equation}
which reduces to \eqref{eq:m_dot_Bondi} for $v_\infty \rightarrow 0$ and a value similar to \eqref{eq:m_dot_HL} in the supersonic limit  $v_\infty \gg a_\infty$ (see, e.g., \cite{Bon52,ShiMTS85,PetSST89}).  Following several previous authors (e.g.~\cite{PetSST89,FonI98b,FonI98a}) we will refer to the approximate accretion rate \eqref{eq:m_dot_can} as the {\em canonical} accretion rate.  

Evidently, \eqref{eq:m_dot_can} predicts that relative motion $v_{\infty} > 0$ reduces the accretion rate; moreover, it predicts that this reduction is monotonic with $v_\infty$, i.e.~that a larger $v_\infty$ will lead to a smaller rate of accretion.  Both Newtonian and relativistic numerical experiments that modeled the accretion of polytropic gases onto moving black holes observed that the accretion rate is indeed often smaller than $\dot M_{\rm sph}$ for $v_\infty > 0$, but usually  larger than $\dot M_{\rm can}$ (see, e.g., the entries for $\dot M/\dot M_{\rm can}$ in Table~1 in \cite{PetSST89} and Table~3 in \cite{FonI98b}). For some values of $\Gamma$, $v_\infty$, and $a_\infty$, however, \cite{PetSST89} and \cite{FonI98b} also found accretion rates {\em greater} than $\dot M_{\rm sph}$.  Newtonian simulations also observed that, for subsonic flows with $v_\infty < a_\infty$, the accretion rate is almost independent of $v_\infty$ (e.g.~\cite{Ruf94,PruGBPR24}) and nearly spherical well inside the effective accretion radius $r_a \simeq GM/(v_{\infty}^2 + a_{\infty}^2)$ (see \cite{Hun71,ShaT83}).

For accretion onto a black hole moving in an extreme-relativistic (ER) fluid with EOS $P=\rho$ (i.e., $a_\infty = c$ and effective adiabatic index $\Gamma = 2$) the accretion has, in fact, been computed analytically, even for rotating black holes (see \cite{PetST88}).  In the
nonspinning case this solution yields a rest-mass accretion rate
\begin{equation} \label{eq:m_dot_ur}
\dot M_0 = \frac{16 \pi G^2 M^2 \rho_{0\infty}  \gamma_\infty}{c^3} ~~~~~\mbox{(ER)}.
\end{equation}
Here $\gamma_\infty = (1 - v_\infty^2/c^2)^{-1/2}$ is the Lorentz factor for the relative motion between the black hole and gas, and accounts for the relativistic increase in the density caused by length contraction as measured in the frame of the black hole.   Note that \eqref{eq:m_dot_ur} indicates a monotonic {\em increase} in the accretion rate with $v_\infty$, rather than a decrease as predicted by \eqref{eq:m_dot_can}.  Exploring these evident distinctions between \eqref{eq:m_dot_can} and \eqref{eq:m_dot_ur} provides one motivation for our study here.

A second motivation arises from the study of small black holes, possibly primordial in nature, that may be caught and swallowed by neutron stars and, once inside, then accrete neutron-star material.  Primordial black holes, originally proposed by \cite{ZelN67,Haw71,CarH74} and others, may well either contribute to, or even make up all, the dark matter content of the Universe (see, e.g., \cite{CarKSY21,CarCGHK24} for reviews).  Several authors have noted that neutron stars would be able to capture some of these black holes and have explored the subsequent co-evolution of the black hole intruder and the neutron star host (see, e.g., \cite{Abretal09,CapPT13,AbrBW18,BraL18,BauS21,AbrBUW22,BauS24c,CaiBK24}, as well as \cite{EasL19,RicBS21b,SchBS21,BauS24d} for numerical simulations).   Akin to the common-envelope evolution of stellar-mass compact objects inside massive stars,\footnote{See, e.g., \cite{MacRR15,CruR20,CruLC26} and references therein for studies of accretion in that context.} this co-evolution hinges on the rate at which the black hole accretes the neutron star material.   Unlike the material accreted in many other astrophysical accretion processes, the neutron star fluid is stiff, with an (effective) adiabatic index $\Gamma > 5/3$, for which the the simple Newtonian expression \eqref{eq:lambda} for the accretion eigenvalue $\lambda$ (generally)  does not even provide a real number.  
As it turns out, accretion of stiff fluids with $\Gamma > 5/3$ has to be 
treated fully relativistically even for {\it spherical} accretion with $v_\infty = 0$ (see, e.g., \cite{Beg78,ChaMS16,RicBS21a}).  In particular, the accretion eigenvalues $\lambda$ are no longer independent of $a_\infty$, which may lead to the cancellation of some powers of $a_\infty$ in \eqref{eq:m_dot_Bondi} and hence change the scaling -- thereby eliminating the basis for finding a correction factor in the canonical accretion rate \eqref{eq:m_dot_can}.   This observation already motivated two of us to consider a canonical accretion rate with an adjusted exponent for the correction factor in \cite{BauS24c}.  In fact, accretion of stiff fluids features some {\em qualitative} differences from accretion of soft fluids; in particular there exists a minimum accretion rate for fluids with $\Gamma > 5/3$ for steady-state flow (see \cite{ChaMS16,RicBS21a}).  These considerations serve as motivation to explore whether {\it nonspherical} accretion onto black holes {\em moving} through stiff fluids also leads to qualitative differences from that for soft fluids -- which is what we probe in this paper. 

A third motivation stems from the need to compute or estimate the drag force exerted on a black hole moving through a fluid -- for example, for the primordial black holes captured inside neutron stars as discussed above.  As discussed, for example, in \cite{PetSST89}, two different physical processes may contribute to this drag: the acquisition of momentum carried by the fluid that is accreted by the black hole, and the exchange of momentum between the fluid and black hole via long-range gravitational forces.  We consider estimates for these {\em capture} and {\em deflection} rates in both the subsonic and supersonic regimes, and compare with drag rates computed from our numerical simulations.

Complementing the work by numerous previous authors, the vast majority of whom focused on Newtonian fluid flow with $\Gamma \leq 5/3$ only (see again Table I in \cite{FogGR05} for a summary) we consider relativistic fluids for EOSs with $\Gamma = 5/3$ and $\Gamma = 2$, for both subsonic and supersonic flow.  Summarized briefly, we find for both values of $\Gamma$ that 
\begin{itemize}
\item for subsonic flow, the accretion rate increases with $v_\infty$ due to relativistic length contraction (which is hence consistent with both (\ref{eq:m_dot_ur}) and the absence
of length contraction in a Newtonian context, {\em cf.} \cite{Ruf94,PruGBPR24}), 

\item for moderately supersonic flow the appearance of shocks may lead to a decrease in the accretion rate, as observed in many previous studies,

\item for strongly relativistic flow, with $v_\infty \lesssim 1$, the accretion rate is dominated by relativistic effects and increases with $v_\infty$, consistent with the findings of \cite{PetST88},

\item we find good agreement between numerical values for the drag rate and our proposed analytical fits for the capture rate alone in the subsonic regime, and a combination of the capture and deflection rate in the supersonic regime.
\end{itemize}

Our paper is organized as follows.  In Sec.~\ref{sec:num} we discuss our numerical approach, including the numerical code, initial and boundary conditions, as well as diagnostics and code checks.  In Sec.~\ref{sec:results} we present numerical results, discuss qualitative features in the fluid flow and accretion rates, compare with previous studies, and propose simple analytical fits.  We briefly summarize in Sec.~\ref{sec:summary}.  For the remainder of this paper we will adopt geometrized units with $G = c = 1$.

%
\section{Numerical Setup}
\label{sec:num}
%

\subsection{Numerical code}
\label{sec:num_code}

For all our simulations we adopt the numerical finite-difference code described in \cite{BauMCM13,MonBM14,BauMM15}.   In particular, the code adopts a spherical polar coordinate system and allows for a logarithmically spaced radial grid, so that the region around a black hole at the coordinate origin can be resolved much more finely than regions far away from the black hole.  While the code does not make any symmetry assumptions, we consider only axisymmetric flow in this paper and therefore use a two-dimensional grid.  

For all our simulations we assume that the total mass in the fluid is much less than the mass of the black hole.  In this limit, we may neglect the self-gravity of the fluid, as well as its effects on the black hole, and therefore perform our simulations in the so-called Cowling approximation -- i.e.~we keep the spacetime geometry fixed and evolve the fluid only.   We describe the Schwarzschild spacetime in maximally sliced trumpet coordinates (see \cite{BauN07}), which are time-independent, can be constructed analytically, and remain regular at the horizon, so that the fluid can cross the horizon without encountering coordinate singularities (see also \cite{MilB17}).   

We solve the equations of relativistic hydrodynamics in spherical polar coordinates using the techniques discussed in \cite{MonBM14,BauMM15}, adopting a Harten-Lax-van-Leer-Einfeldt approximate Riemann solver \cite{HarLL83,Ein88} together with a simple monotonized central-difference limiter reconstruction scheme \cite{Van77}.  While the initial data satisfy a polytropic EOS of the form (\ref{eq:polytrope}), we employ a Gamma-law EOS
\begin{equation} \label{eq:gamma_law}
P = (\Gamma - 1) \rho_0 \epsilon    
\end{equation}
during the evolution.  In (\ref{eq:gamma_law}), $\epsilon$ is the specific energy density, and we note that Eq.~(\ref{eq:polytrope}) follows from Eq.~(\ref{eq:gamma_law}) for isentropic flow.  For a shocked fluid, on the other hand, $K$ in (\ref{eq:polytrope}) increases, so that (\ref{eq:gamma_law}) is no longer consistent with the polytropic form (\ref{eq:polytrope}) with constant $K$. 

In order to prevent numerical artifacts close to the black-hole singularity at $r = 0$ we excise the evolution of the hydrodynamical variables on the innermost three gridpoints in each angular direction well inside the horizon.

\subsection{Initial conditions}
\label{sec:num_indata}

\begin{figure*}[t]
    \centering
    \includegraphics[width=.98\textwidth]{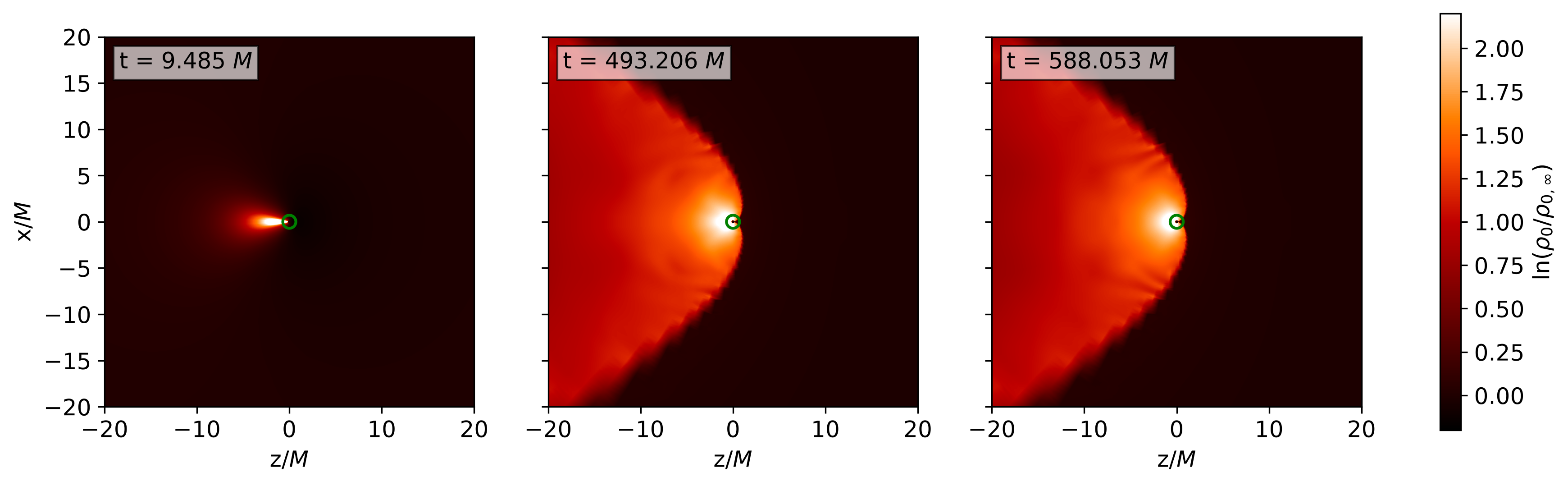}
    \caption{Rest-mass density pseudocolor plots for the evolution of constant-density initial data for $\Gamma = 2$, $a_\infty = a_{\rm out} = 0.1$, and $v_\infty = 0.5$ (see Table \ref{tab:GammaTwo} for more details).  The horizontal axis ($z$) represents the axis of symmetry, with the fluid flow from right to left, and vertical axis ($x$) shown in the equatorial plane.   The green circles represent the black hole event horizon.  The left panel shows a contour at an early time $t = 9.48 M$, during the transition from the constant-density initial data to the steady-state profile, while the middle and right panels show density contours at times $t = 493.21 M$ and $588.05 M$, respectively.  There is very little difference between the profiles at the the latter two  late times, indicating that the flow has settled down into a stationary solution.}
    \label{fig:settlingpanels}
\end{figure*}

We solve for the fluid flow in the rest frame of the black hole, which we place at the origin of our spherical grid. We consider two types of initial data for our simulations.  One type of data,  which we refer to as {\em Bondi-type} initial data (labeled BON in Tables \ref{tab:GammaFiveThirds} and \ref{tab:GammaTwo}), is constructed from the analytical Bondi solution for spherical accretion of a fluid at rest at large separations from the black hole of mass $M$, for which the accretion rate is given by \eqref{eq:m_dot_Bondi} (see \cite{Bon52,Mic72,ShaT83}).  Given choices for the adiabatic index $\Gamma$ and the asymptotic sound speed $a_\infty$ we construct profiles of the fluid density $\rho_0$ and four-velocity $u^a$ as a function of radius $r$ as discussed in \cite{RicBS21a}.  We transform these solutions into the maximally-sliced trumpet coordinates of \cite{BauN07} and compute the spatial velocity field
\begin{equation} \label{eq:vi}
v^i \equiv \frac{1}{\gamma} \gamma^i_{~a} u^a 
= \frac{1}{\alpha} \left(\frac{u^i}{u^t} + \beta^i \right)
\end{equation}
as discussed in \cite{MilB17}.  Here $\gamma_{ij} = g_{ab} + n_a n_b$ is the spatial metric, i.e.~the projection of the spacetime metric $g_{ab}$ onto a spatial slice with normal vector $n_a = (\alpha, 0, 0, 0)$, $\alpha$ is the lapse function, $\beta^i$ the shift vector, and $\gamma \equiv - n_a u^a = \alpha u^t$ is the Lorentz factor between a normal observer (whose four-velocity is the normal vector $n^a$) and an observer comoving with the fluid.\footnote{In the context of relativistic hydrodynamics, this Lorentz factor is often denoted by $W$.  We adopt the symbol $\gamma$ here, but caution that it should not be confused with the determinant of the spatial metric $\gamma_{ij}$.}  In terms of the spatial velocity $v^i$ as defined in (\ref{eq:vi}), the latter can also be written in the ``usual" form  
\begin{equation}
\gamma = \left( 1 - v_i v^i \right)^{-1/2}.
\end{equation}

In order to endow the fluid with a relative asymptotic speed $v_\infty$, we superimpose on the spherical Bondi velocity field $v_{\rm sph}^r$ a velocity field aligned with the $z$-axis and with speed $v_\infty$ at infinity.  Specifically, we choose 
\begin{subequations} \label{eq:vel_profile}
\begin{align} 
v_{\rm init}^r & = v_{\rm sph}^r - v_\infty\tanh{\left(\frac{r}{\sigma M}\right)}\cos{(\theta)} \\
v_{\rm init}^\theta & =  v_\infty\tanh{\left(\frac{r}{\sigma M}\right)}\sin{(\theta)}
\end{align}
\end{subequations}
Here the $\tanh$ terms approach unity as $r \gg \sigma M$, resulting in the desired fluid flow at large separations from the black hole, but prevent $v_{\rm init}$ from artificially exceeding the speed of light close to the origin at $r = 0$.  In practice we choose the dimensionless parameter $\sigma = 5$, except for some simulations with $v_\infty \lesssim a_{\infty}$, in which case we use slightly larger values of $\sigma$.

As a second type of initial data we consider {\em constant-density} initial data (labeled CD in Tables \ref{tab:GammaFiveThirds} and \ref{tab:GammaTwo}).  In this case we simply set all hydrodynamical variables to their asymptotic values for given values of $\Gamma$ and $a_\infty$ and use \eqref{eq:vel_profile} with $v_{\rm sph}^r = 0$. 

We emphasize that, as we evolve to steady-state, the initial conditions we specify in the interior can be quite arbitrary, as they will be washed out as the simulation evolves and the profiles settle down (see Figs.~\ref{fig:settlingpanels} and \ref{fig:initdataconv} as well as Sec.~\ref{sec:num_evolve} below). The outer boundary conditions, by contrast, determine the nature of the solution.   

\subsection{Boundary conditions}
\label{sec:num_bc}

Our numerical finite-difference grid extends to an outer boundary at $r = r_{\rm out}$, which should be chosen to be large in comparison to the accretion radius $r_a$.  The accretion radius corresponds to the critical point at which the numerator and denominator in the equation governing the spherical accretion flow vanish simultaneously (the transonic radius in the Newtonian limit; see, e.g., \cite{ShaT83}, for details).  In the limit of $v_\infty \rightarrow 0$, $r_a$ is often estimated, up to a constant factor of order unity, from   
\begin{equation} \label{r_a_wrong}
r_a \approx \frac{5 - 3 \Gamma}{4} \frac{M}{a_\infty^2} \quad (\Gamma < 5/3)
\end{equation} 
(see, e.g., Eq.~(2.43) in \cite{PetSST89}), even though the scaling $r_a \propto a_\infty^{-2}$ holds only for $\Gamma < 5/3$.  For $\Gamma = 5/3$, and to leading order in $a_\infty$, we instead have
\begin{equation}
r_a \approx \frac{3}{4} \frac{M}{a_\infty} \quad (\Gamma = 5/3,~a_\infty \ll 1)
\end{equation}
(see Exercise G.1 in \cite{ShaT83} and Section 3.1.1 in \cite{RicBS21a}), while for $\Gamma > 5/3$ the accretion radius approaches a constant value in the limit $a_\infty \rightarrow 0$,
\begin{equation}
r_a = \frac{3 \Gamma - 5/3 - (12 \Gamma - 11)^{1/2}/2}{2 \Gamma - 7/3 - (12 \Gamma - 11)^{1/2}/3} M \quad (\Gamma > 5/3,~a_\infty \ll 1)
\end{equation}
(see Section 3.1.2 in \cite{RicBS21a} and the Erratum that corrects their Eq.~(43)).  For stiff EOSs, with $\Gamma \geq 5/3$, and for small values of $a_\infty$, $r_a$ is therefore significantly smaller than the estimate (\ref{r_a_wrong}) would suggest.

In addition to grid points inside $r_{\rm out}$, which we refer to as interior grid points, the finite-differencing also requires a few exterior grid points outside $r_{\rm out}$ that have to be filled using the boundary conditions.  For the purposes of our simulations here we distinguish between {\em upstream} and {\em downstream} conditions.   

For upstream exterior grid points, namely those for $0 < \theta < \pi/2$, we keep all fluid variables fixed to those specified in the initial data.  This choice ensures that the upstream boundary continues to provide inflowing fluid throughout the simulation. 

For downstream exterior grid points, for $\pi/2 < \theta < \pi$, on the other hand, we copy, for each value of $\theta$, values of the fluid variables from the outermost interior grid point (just inside $r_{\rm out}$) to the exterior grid points.  We have verified that there is no discontinuity across the upstream-downstream boundary at $\theta = \pi/2$ in the steady-state solution.

We generally adopt Bondi-type initial data for subsonic simulations, and constant-density data for supersonic simulations. This choice is primarily motivated by a subtle but important difference that results from the fact that the boundary conditions are imposed at a finite value of $r_{\rm out}$.

The Bondi-type initial data feature a radial profile. In this case, the sound speed takes a value $a_{\rm out}$ at $r_{\rm out}$ that is generally greater than its asymptotic value $a_\infty$ corresponding to $r \rightarrow \infty$.  Our dynamical evolutions verify that for {\em subsonic} flows ($v_\infty < a_\infty$), the upstream profile remains close to the Bondi profile, meaning that the correct value of $a_{\rm out}$ imposed at the numerical boundary should be greater than $a_{\infty}$ -- as is the case for Bondi-type initial data.

By contrast, the fluid variables at $r_{\rm out}$ for constant-density initial data are, by construction, the same as those at an infinite separation $r \rightarrow \infty$ from the black hole. For {\em supersonic} flows, the upstream fluid cannot be affected by the downstream black hole, meaning that $a_{\rm out}$ should indeed equal $a_{\infty}$ -- as is the case for constant-density initial data. 

We report the ratio $a_{\rm out}/a_{\infty}$ for all our simulations in Tables \ref{tab:GammaFiveThirds} and \ref{tab:GammaTwo}, and emphasize that these ratios approach unity, of course, as $r_{\rm out} \rightarrow \infty$.

For most cases that we considered in Sec.~\ref{sec:results} below we found that it was sufficient to place the outer boundary at $r_{\rm out} = 200 M$, and to adopt a modest grid resolution with $N_r = 248$ (logarithmically spaced) radial and $N_\theta = 36$ angular grid points, to obtain steady-state accretion rates to within about a percent accuracy. In some cases, however, a higher grid resolution was needed in order to reduce noisy oscillations in the accretion rates at late times, and for some subsonic cases we increased the value of $r_{\rm out}$ in order to reduce effects of the downstream boundary condition.  We therefore list the values of $N_r$, $N_\theta$, and $r_{\rm out}$ for each run listed in Tables \ref{tab:GammaFiveThirds} and \ref{tab:GammaTwo}. 

\subsection{Evolution}
\label{sec:num_evolve}

We evolve the initial data described in Sec.~\ref{sec:num_indata}, subject to the boundary conditions discussed in Sec.~\ref{sec:num_bc}, until they have settled down into an approximately stationary flow.  As an example we show in Fig.~\ref{fig:settlingpanels} the evolution for a stiff EOS ($\Gamma = 2$) with supersonic flow, starting with constant density initial data.  In the left panel we show a density contour plot during the early transient phase, while the flow settles down, and in the middle and right panels we show the stationary density profile at later times.  We did not encounter instabilities as those discussed in \cite{FogGR05}, including the so-called flip-flop instability (see also \cite{DoeZR11}).

 \begin{figure}[t]
    \centering
    \includegraphics[width=0.48\textwidth]{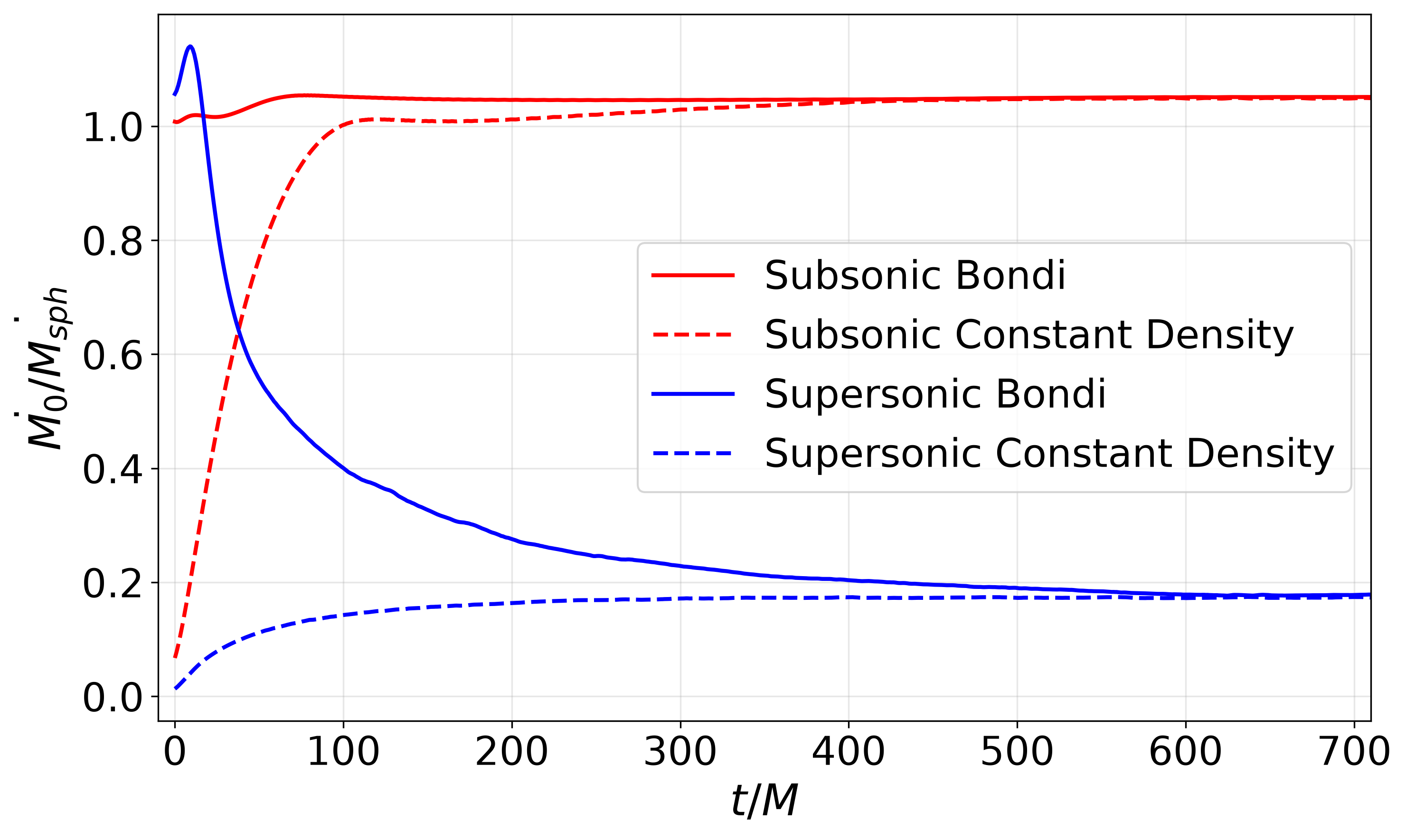}
    \caption{
    The rest-mass accretion rate $\dot M_0$ as a function of time $t$ for a stiff fluid with $\Gamma = 2$.  We show examples for both subsonic flow (blue lines, with $a_{\rm out} = 0.31$ and $v_\infty = 0.2$) and supersonic flow (red lines, with $a_{\rm out} = 0.12$ and $v_\infty = 0.5$). For each flow we show evolutions using both Bondi-type initial data (solid lines) and constant-density initial data (dotted lines).   Note that Bondi-type initial data settle down more rapidly than the constant-density initial data for subsonic flow, but the reverse is true for supersonic flow.  Also note that both types of initial data settle down to very similar accretion rates at late times, as expected.}
    \label{fig:initdataconv}
\end{figure} 

In Fig.~\ref{fig:initdataconv} we show the rest-mass accretion rate $\dot M_0$ (see Sec.~\ref{sec:num_restmass} below) for a stiff EOS for both Bondi-type and constant-density initial data, and for both subsonic and supersonic flow.  As expected, both types of initial data lead to very similar asymptotic accretion rates once the evolution has settled down to steady-state.  Note, however, that Bondi-type initial data tend to settle down faster than constant-density initial data for subsonic flow, while the converse is true for supersonic flow, adding a secondary motivation for our choice of initial data discussed in Sec.~\ref{sec:num_bc}.

\subsection{Diagnostics and code checks}
\label{sec:num_diagnostics}

During the evolution we monitor several different diagnostics, including accretion rates as well as the drag rate.

\subsubsection{Rest-mass accretion rate}
\label{sec:num_restmass}

The rate at which the black hole accretes rest mass is related to the fluid flow across the black hole horizon ${\mathcal H}$ by 
\begin{equation} \label{eq:accretionrate}
\dot M_0 = - \oint_{\mathcal H} \alpha \rho_0 u^i dS_i,
\end{equation}
where $dS_i$ is the outward-oriented surface element on $\mathcal H$ (see, e.g., Eq.~(33) in \cite{BauS24d}).  In steady-state, conservation of rest mass guarantees that the accretion rate can be measured from an integral over any surface, not just the horizon $\mathcal H$.

\begin{figure}[t]
    \centering
    \includegraphics[width = 0.48 \textwidth]{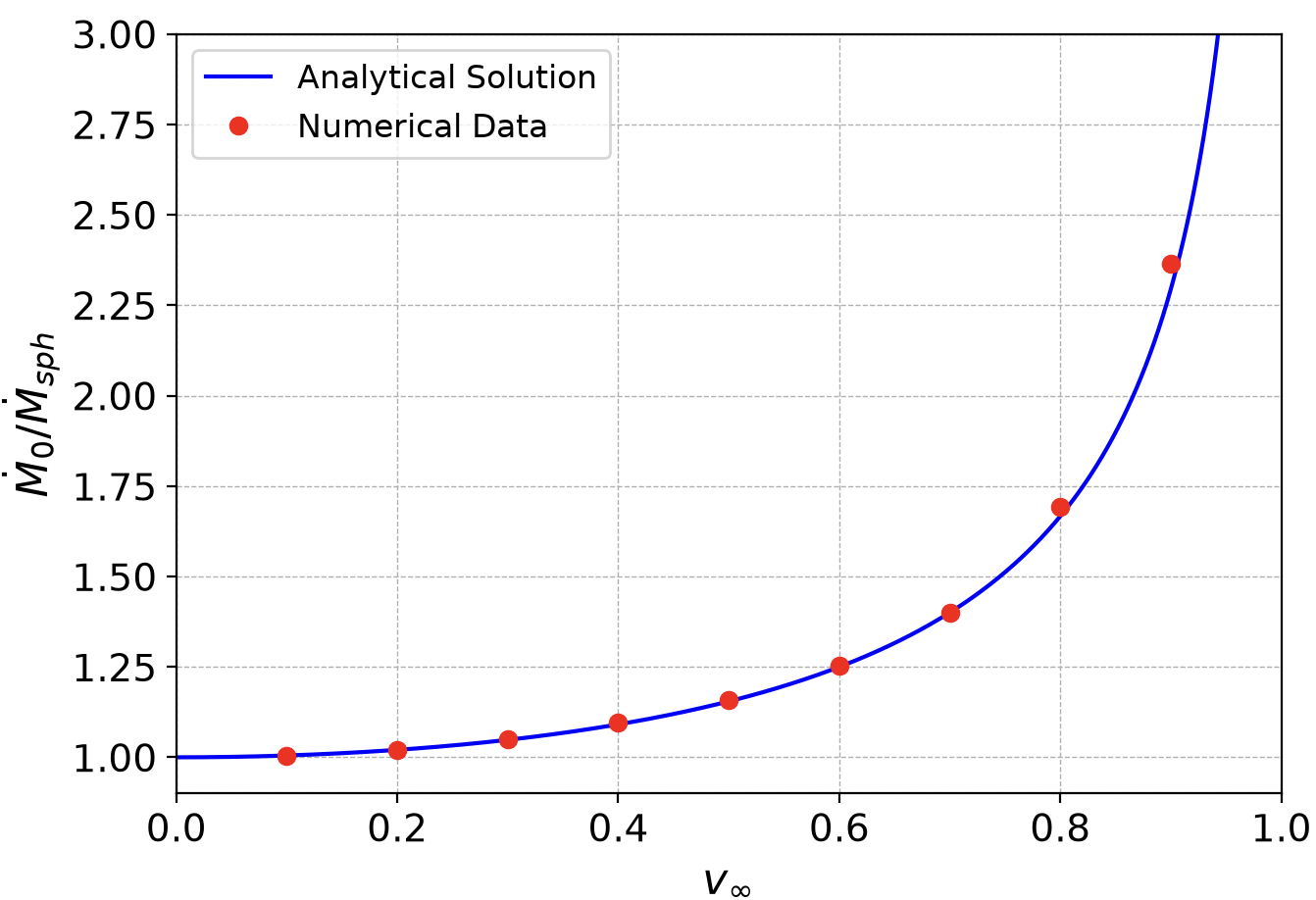}
    \caption{Rest-mass accretion rates $\dot M_0$ (see Eq.~(\ref{eq:accretionrate})) versus $v_\infty$ for ultrarelativistic $P=\rho$ fluids with $\Gamma = 2$.  The solid line shows the analytical values (\ref{eq:m_dot_ur}) for the analytical solution of \cite{PetST88} (i.e.~for $a_\infty = 1$), while the dots show our numerical results for $a_\infty = 0.9999$.  }
    \label{fig:ultrarelcomp}
\end{figure}

\begin{figure}[t]
    \centering
    \includegraphics[width=0.48\textwidth]{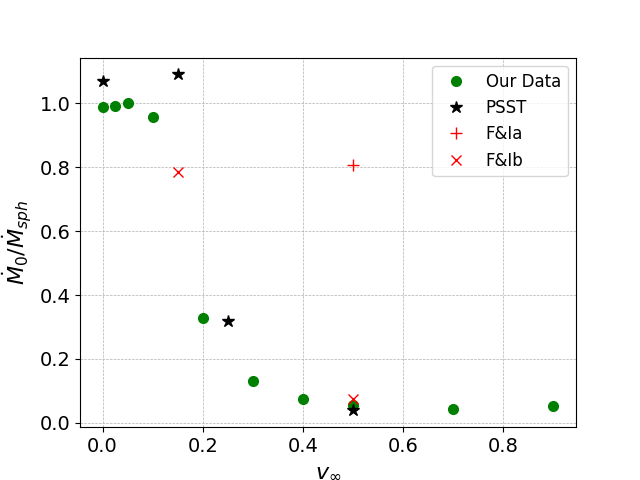}
    \caption{Comparison of our results for the rest-mass accretion rate $\dot M_0$ for $\Gamma = 5/3$ and $a_\infty = 0.1$ with those of Petrich {\it et.al.} (PSST, \cite{PetSST89}) as well as Font and Iba\~nez (F\&Ia, \cite{FonI98a}, and F\&Ib, \cite{FonI98b}). }
    \label{fig:g53a01_compare}
\end{figure}

As an immediate test we confirm that, for $v_\infty = 0$, our simulations yield the expected accretion rate \eqref{eq:m_dot_Bondi} for spherical Bondi accretion to within less than one percent (see the entries for $v_\infty = 0$ in Tables \ref{tab:GammaFiveThirds} and \ref{tab:GammaTwo}). As another code check we compare with the analytical accretion rates \eqref{eq:m_dot_ur} of \cite{PetST88}, which hold for an ultrarelativistic fluid with $\Gamma = 2$ and $a_\infty = 1$.  In our numerical simulations we adopt $a_\infty = 0.9999$ in order to avoid numerical issues, but, as shown in Fig.~\ref{fig:ultrarelcomp}, our accretion rates are nevertheless very close to the analytical solution.  

Finally, we compare our results for the rest-mass accretion rate $\dot M_0$ with those of \cite{PetSST89,FonI98a,FonI98b} in Fig.~\ref{fig:g53a01_compare} for $\Gamma = 5/3$.  While the value for $\dot M_0 / \dot M_{\rm sph}$ for $v_\infty = 0.5$ reported in \cite{FonI98b} agrees rather well with both our results and those of \cite{PetSST89}, the value reported in \cite{FonI98a} is larger by almost exactly a factor of ten, suggesting that this discrepancy might be the result of a typo.  For completeness we include both values in Fig.~\ref{fig:g53a01_compare}.

\subsubsection{Mass-energy accretion rate}

As discussed in Sec.~\ref{sec:intro}, the rest-mass accretion rate $\dot M_0$ is different from the mass-energy accretion rate $\dot M$.  The latter accounts for the accretion of other forms of energy in addition to rest-mass and leads to the increase of the black hole's gravitational mass (see also the discussion in \cite{AshK03,AguST21,AguTSL21,RicBS21b} as well as the Appendix in \cite{BauS24d}). For steady-state, spherical fluid accretion, the mass-energy accretion rate $\dot M$ can be estimated from (\ref{eq:M_grav_dot}), i.e.~by multiplying the rest-mass accretion rate $\dot M_0$ with the value of the asymptotic fluid enthalpy 
\begin{equation}
    h_\infty = \frac{\Gamma - 1}{\Gamma - 1 - a_\infty^2}.
\end{equation}
Still accounting for the fluid only, the total mass-energy accretion rate of a black hole moving through a gas is given by
\begin{equation} \label{eq:M_grav_dot_v}
\dot M = h_\infty \gamma_\infty \dot M_0
\end{equation}
(see Eq.~(2.31) in \cite{PetSST89}), where the Lorentz factor $\gamma_\infty$ accounts for the fluid's kinetic energy.  In Tables \ref{tab:GammaFiveThirds} and \ref{tab:GammaTwo} below we provide numerical results for $\dot M$ computed from (\ref{eq:M_grav_dot_v}).

\subsubsection{Drag rate}

\begin{figure}[t]
    \centering
    \includegraphics[width=0.48\textwidth]{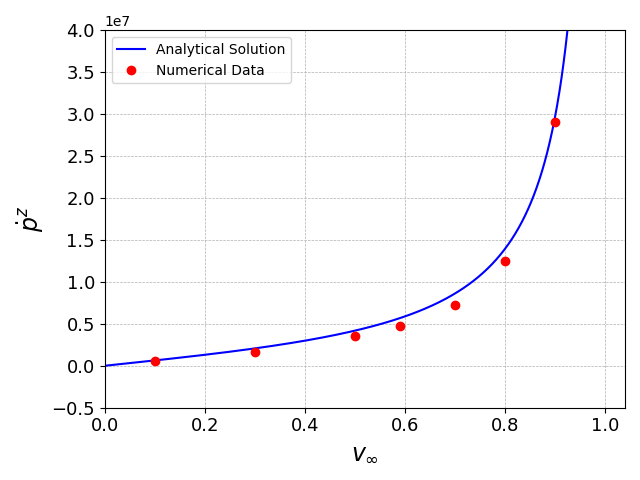}
    \caption{Same as Fig.~\ref{fig:ultrarelcomp}, but for the drag rate $\dot{p}^z$.  Numerical results are computed from Eq.~(\ref{eq:dragrate}), while the analytical results are given by Eq.~(\ref{eq:p_dot_ur}).}
    \label{fig:ur_drag}
\end{figure}

We also compute the drag rate, i.e.~the rate at which the black hole acquires momentum $p^z$, from
\begin{equation} \label{eq:dragrate}
    \dot{p}^z \approx - \oint {T^{zj} \sqrt{-g}dS_j}
\end{equation}
(see, e.g., Eq.~(2.32) in \cite{PetSST89}).  We note that (\ref{eq:dragrate}) is exact only in the presence of a Killing vector that is aligned with $p^z$ (see, e.g., Prob.~10.11 in \cite{LigPPT75}) -- which, for the Schwarzschild spacetimes considered here, is true only asymptotically.\footnote{An exact treatment involves the introduction of a gravitational pseudotensor, which accounts for corrections to the balance law (\ref{eq:dragrate}) in the absence of an exact Killing vector; see \cite{Clo21} for a discussion and \cite{TraVCHBFH23,WanCCM26}, for example, for applications.}  We therefore evaluate the surface integral on the right-hand side of (\ref{eq:dragrate}) at large distances from the black hole (see Appendix  \ref{sec:Appendix_A_int} for more details).

In Appendix \ref{sec:Appendix_A_analytical} we provide an analytical result for the drag rate in the accretion of the ultrarelativistic fluid considered by \cite{PetST88} (see Eq.~(\ref{eq:p_dot_ur})).  In Fig.~\ref{fig:ur_drag} we compare our numerical results with this analytical solution.  While the agreement is not quite as good as that for the rest-mass accretion rates shown in Fig.~\ref{fig:ultrarelcomp}, this is not too surprising given the inherent problems associated with evaluating the drag rate (see the discussion in Appendix  \ref{sec:Appendix_A_int}).

%
\section{Results}
\label{sec:results}
%

\begin{table*}[t]
    \centering
    \begin{tabular}{cccccccccc}
          \toprule
          $a_{\infty}$ & $v_{\infty}$ & $\mathcal{M}_{\infty}$ & $\dot{M_0}/\dot{M}_{\mathrm{sph}}$ & $\dot{M}/\dot{M}_0$ & $\dot{p}^z/(M^2\rho_\infty{v_\infty})$ & ${a_{\rm out}}/{a_{\infty}}$ & $r_{\rm out}/M$ & $(N_r, N_\theta)$ & ID \\
          \hline
          0.1 & 0.0 & 0.00 & 1.00 & 1.015 & ... & 1.07 & 400 & (376, 2) & BON \\
          ... & 0.025 & 0.25 & 1.00 & 1.016 & 3300 & 1.07 & 400 & (376, 36) & BON \\
          ... & 0.05 & 0.50 & 1.01 & 1.016 & 3300 & 1.07 & 400 & (376, 36) & BON \\
          ... & 0.1 & 1.00 & 0.974 & 1.020 & 4200 & 1.07 & 400 & (376, 36) & BON \\
          ... & 0.2 & 2.00 & 0.325 & 1.036 & 2300 & 1.00 & 200 & (248, 36) & CD \\
          ... & 0.3 & 3.00 & 0.133 & 1.064 & 1400 & 1.00 & 200 & (248, 36) & CD \\
          ... & 0.4 & 4.00 & 0.0763 & 1.108 & 890 & 1.00 & 200 & (248, 36) & CD \\
          ... & 0.5 & 5.00 & 0.0553 & 1.172 & 646 & 1.00 & 200 & (248, 36) & CD \\
          ... & 0.7 & 7.00 & 0.0427 & 1.422 & 521 & 1.00 & 200 & (248, 36) & CD \\
          ... & 0.9 & 9.00 & 0.0522 & 2.329 & 1000 & 1.00 & 200 & (248, 72) & CD \\
          \hline
          0.3 & 0.0 & 0.00 & 0.998 & 1.156 & ... & 1.02 & 200 & (248, 2) & BON \\
          ... & 0.1 & 0.33 & 1.02 & 1.162 & 410 & 1.02 & 200 & (248, 36) & BON \\ 
          ... & 0.2 & 0.67 & 1.08 & 1.180 & 396 & 1.02 & 200 & (248, 36) & BON \\
          ... & 0.3 & 1.00 & 1.16 & 1.212 & 650 & 1.00 & 200 & (248, 36) & CD \\
          ... & 0.4 & 1.33 & 1.02 & 1.261 & 750 & 1.00 & 200 & (248, 36) & CD \\
          ... & 0.5 & 1.67 & 0.736 & 1.335 & 590 & 1.00 & 200 & (248, 36) & CD\\
          ... & 0.7 & 2.33 & 0.559 & 1.619 & 500 & 1.00 & 200 & (248, 36) & CD \\
          ... & 0.9 & 3.00 & 0.700 & 2.652 & 1000 & 1.00 & 200 & (248, 72) & CD \\
          \hline
          0.5 & 0.0 & 0.00 & 0.997 & 1.600 & ... & 1.00 & 200 & (248, 2) & BON \\
          ... & 0.1 & 0.20 & 1.01 & 1.608 & 152 & 1.00 & 200 & (248, 36) & BON \\
          ... & 0.2 & 0.40 & 1.05 & 1.633 & 119 & 1.00 & 200 & (248, 36) & BON \\
          ... & 0.3 & 0.60 & 1.12 & 1.677 & 160 & 1.00 & 200 & (248, 36) & BON \\
          ... & 0.4 & 0.80 & 1.23 & 1.746 & 191 & 1.00 & 200 & (248, 36) & BON \\
          ... & 0.5 & 1.00 & 1.37 & 1.848 & 310 & 1.00 & 200 & (248, 36) & CD \\
          ... & 0.6 & 1.20 & 1.50 & 2.000 & 440 & 1.00 & 200 & (248, 36) & CD \\
          ... & 0.7 & 1.40 & 1.47 & 2.240 & 540 & 1.00 & 200 & (248, 36) & CD \\
          ... & 0.9 & 1.80 & 1.87 & 3.671 & 1040 & 1.00 & 200 & (248, 36) & CD \\
          \hline
          0.81 & 0.0 & 0.00 & 0.997 & 63.09 & ... & 1.00 & 200 & (248, 2) & BON \\
          ... & 0.1 & 0.12 & 1.01 & 63.41 & 110 & 1.00 & 200 & (248, 36) & BON \\
          ... & 0.2 & 0.25 & 1.03 & 64.39 & 95 & 1.00 & 200 & (248, 36) & BON \\
          ... & 0.3 & 0.37 & 1.08 & 66.14 & 102 & 1.00 & 200 & (248, 36) & BON \\
          ... & 0.4 & 0.49 & 1.14 & 68.84 & 119 & 1.00 & 200 & (248, 36) & BON \\
          ... & 0.49 & 0.60 & 1.23 & 72.38 & 130 & 1.00 & 200 & (248, 36) & BON \\
          ... & 0.6 & 0.74 & 1.40 & 78.86 & 180 & 1.00 & 200 & (248, 36) & BON \\
          ... & 0.7 & 0.86 & 1.66 & 88.35 & 240 & 1.00 & 200 & (248, 36) & BON \\
          ... & 0.81 & 1.00 & 2.16 & 107.6 & 352 & 1.00 & 200 & (248, 36) & CD \\
          ... & 0.9 & 1.11 & 3.24 & 144.7 & 1060 & 1.00 & 200 & (248, 36) & CD \\
          \hline
    \end{tabular}
    \caption{Numerical results for $\Gamma = 5/3$.  For each value of the asymptotic sound speed $a_\infty$ and the relative speed $v_\infty$, we list the corresponding Mach number ${\mathcal M}_\infty = v_\infty / a_\infty$, the rest-mass accretion rate $\dot M_0$ in units of the spherical (rest-mass) accretion rate $\dot M_{\rm sph}$ (see (\ref{eq:m_dot_Bondi})), the total mass-energy accretion rate $\dot M$ in units for $\dot M_0$ (computed from (\ref{eq:M_grav_dot_v})), and the momentum accretion rate $\dot{p}^z$ in units of $M^2 \rho_\infty v_\infty$.  We also provide the ratio between the sound speed at the outer boundary and its asymptotic value, the location of the outer boundary $r_{\rm out}$, the number of radial and angular grid points $N_r$ and $N_\theta$, and the type of initial data (ID), where BON refers to Bondi-type data and CD to constant-density data. }
    \label{tab:GammaFiveThirds}
\end{table*}

\begin{table*}[t]
    \centering
    \begin{tabular}{cccccccccc}
          \toprule
          $a_{\infty}$ & $v_{\infty}$ & $\mathcal{M}_{\infty}$ & $\dot{M_0}/\dot{M}_{\mathrm{sph}}$& $\dot{M}/\dot{M}_0$ & $\dot{p}^z/(M^2\rho_\infty{v_\infty})$ & ${a_{\rm out}}/{a_{\infty}}$ & $r_{\rm out}/M$ & $(N_r, N_\theta)$ & ID \\
          \hline
          0.1 & 0.0 & 0.00 & 0.998 & 1.010 & ... & 1.11 & 400 & (376, 2) & BON \\
          ... & 0.025 & 0.25 & 1.00 & 1.010 & 1970 & 1.11 & 400 & (376, 36) & BON \\
          ... & 0.05 & 0.50 & 1.00 & 1.011 & 1380 & 1.11 & 400 & (376, 36) & BON \\
          ... & 0.1 & 1.00 & 0.970 & 1.015 & 1080 & 1.11 & 400 & (376, 36) & BON \\
          ... & 0.15 & 1.50 & 0.583 & 1.022 & 1970 & 1.00 & 200 & (248, 36) & CD \\
          ... & 0.2 & 2.00 & 0.429 & 1.031 & 1580 & 1.00 & 200 & (248, 36) & CD \\
          ... & 0.3 & 3.00 & 0.235 & 1.059 & 1050 & 1.00 & 200 & (248, 36) & CD \\
          ... & 0.4 & 4.00 & 0.164 & 1.102 & 640 & 1.00 & 200 & (248, 36) & CD \\
          ... & 0.5 & 5.00 & 0.113 & 1.166 & 470 & 1.00 & 200 & (248, 36) & CD \\
          ... & 0.7 & 7.00 & 0.0920 & 1.414 & 420 & 1.00 & 200 & (248, 36) & CD \\
          ... & 0.9 & 9.00 & 0.123 & 2.317 & 880 & 1.00 & 200 & (248, 36) & CD \\
          \hline
          0.3 & 0.0 & 0.00 & 0.996 & 1.099 & ... & 1.02 & 200 & (248, 2) & BON \\
          ... & 0.1 & 0.33 & 1.01 & 1.104 & 239 & 1.02 & 200 & (248, 36) & BON \\
          ... & 0.2 & 0.67 & 1.04 & 1.122 & 270 & 1.02 & 200 & (248, 36) & BON \\
          ... & 0.3 & 1.00 & 1.10 & 1.152 & 390 & 1.00 & 200 & (248, 36) & CD \\
          ... & 0.4 & 1.33 & 0.996 & 1.199 & 530 & 1.00 & 200 & (248, 36) & CD \\
          ... & 0.5 & 1.67 & 0.855 & 1.269 & 480 & 1.00 & 200 & (248, 36) & CD \\
          ... & 0.7 & 2.33 & 0.698 & 1.539 & 427 & 1.00 & 200 & (248, 36) & CD \\
          ... & 0.9 & 3.00 & 0.894 & 2.521 & 835 & 1.00 & 200 & (248, 36) & CD \\
          \hline
          0.5 & 0.0 & 0.00 & 0.998 & 1.333 & ... & 1.01 & 200 & (248, 2) & BON \\
          ... & 0.1 & 0.20 & 1.01 & 1.340 & 119 & 1.01 & 200 & (248, 36) & BON \\
          ... & 0.3 & 0.60 & 1.09 & 1.398 & 120 & 1.01 & 200 & (248, 36) & BON \\
          ... & 0.4 & 0.80 & 1.16 & 1.455 & 145 & 1.01 & 200 & (248, 36) & BON \\
          ... & 0.5 & 1.00 & 1.26 & 1.540 & 190 & 1.00 & 200 & (248, 36) & CD \\
          ... & 0.6 & 1.20 & 1.35 & 1.667 & 340 & 1.00 & 200 & (248, 36) & CD \\
          ... & 0.7 & 1.40 & 1.40 & 1.867 & 400 & 1.00 & 200 & (248, 36) & CD \\
          ... & 0.9 & 1.80 & 1.84 & 3.059 & 860 & 1.00 & 200 & (248, 36) & CD \\
          \hline
          0.7 & 0.0 & 0.00 & 0.997 & 1.961 & ... & 1.00 & 200 & (248, 2) & BON \\
          ... & 0.1 & 0.14 & 1.01 & 1.971 & 90 & 1.00 & 200 & (248, 36) & BON \\
          ... & 0.3 & 0.43 & 1.07 & 2.055 & 94 & 1.00 & 200 & (248, 36) & BON \\
          ... & 0.5 & 0.71 & 1.21 & 2.264 & 117 & 1.00 & 200 & (248, 36) & BON \\
          ... & 0.6 & 0.86 & 1.37 & 2.451 & 140 & 1.00 & 200 & (248, 36) & BON \\
          ... & 0.7 & 1.00 & 1.53 & 2.756 & 180 & 1.00 & 200 & (248, 36) & CD \\
          ... & 0.8 & 1.14 & 1.86 & 3.268 & 440 & 1.00 & 200 & (248, 36) & CD \\
          ... & 0.9 & 1.29 & 2.47 & 4.498 & 750 & 1.00 & 200 & (248, 36) & CD \\
          \hline
          0.9999 & 0.0 & 0.00 & 0.997 & 5000 & ... & 1.00 & 200 & (248, 2) & BON \\
          ... & 0.1 & 0.10 & 1.01 & 5025 & 92 & 1.00 & 200 & (248, 36) & BON \\
          ... & 0.3 & 0.30 & 1.06 & 5242 & 91 & 1.00 & 200 & (248, 36) & BON \\
          ... & 0.5 & 0.52 & 1.17 & 5774 & 112 & 1.00 & 200 & (248, 36) & BON \\
          ... & 0.59 & 0.60 & 1.26 & 6193 & 129 & 1.00 & 200 & (248, 36) & BON \\
          ... & 0.7 & 0.71 & 1.44 & 7002 & 167 & 1.00 & 200 & (248, 36) & BON \\
          ... & 0.8 & 0.81 & 1.73 & 8334 & 250 & 1.00 & 200 & (248, 36) & BON \\
          ... & 0.9 & 0.91 & 2.41 & 11470 & 516 & 1.00 & 200 & (248, 72) & BON \\
          \hline
    \end{tabular}
    \caption{Same as Table \ref{tab:GammaFiveThirds} but for $\Gamma = 2$. }
    \label{tab:GammaTwo}
\end{table*}

\subsection{Numerical data}
\label{sec:res_data}

We compute the accretion and drag rates of a non-rotating black hole moving through fluids with adiabatic exponents $\Gamma = 5/3$ and $\Gamma = 2$ for different values of relative speeds $v_\infty$ and asymptotic sound speeds $a_\infty$.  We report our numerical results in Tables \ref{tab:GammaFiveThirds} and \ref{tab:GammaTwo}, where in addition to numerical details for the different simulations, we list the rest-mass accretion rates $\dot M_0$, the total mass-energy accretion rates, as well as the drag rates $\dot p^z$.  

In the tables, $\dot M_{\rm sph}$ is given by Eq.~(\ref{eq:m_dot_Bondi}), with $\lambda$ evaluated relativistically from Eqs.~(12) and (16) in \cite{RicBS21a} (called $\lambda_{\rm GR}$ there).  For $a_\infty \ll 1$, values of $\lambda$ can be approximated using an expansion as discussed in Sect.~3.1 of \cite{RicBS21a}.  For $\Gamma \leq 5/3$, this expansion yields Eq.~(\ref{eq:lambda}) to leading order and, in particular, $\lambda = 1/4$ for $\Gamma = 5/3$.  For $\Gamma > 5/3$, values of $\lambda$ are given by Eqs.~(44) and (45) in \cite{RicBS21a}; specifically, for $\Gamma = 2$ we have $\lambda = 1.49 \, a_\infty$ in this limit.  

By contrast, for $1 - a_\infty \ll 1$ (see Sect.~3.1 in \cite{RicBS21a}) and $\Gamma = 2$ we have $\lambda \simeq 4 a_\infty^2 + (5/16)(1 - a_\infty^2)^2$ to leading order (see their Eq.~(55)).  

\begin{figure*}[t]
        \centering
        \includegraphics[width= 0.48 \textwidth]{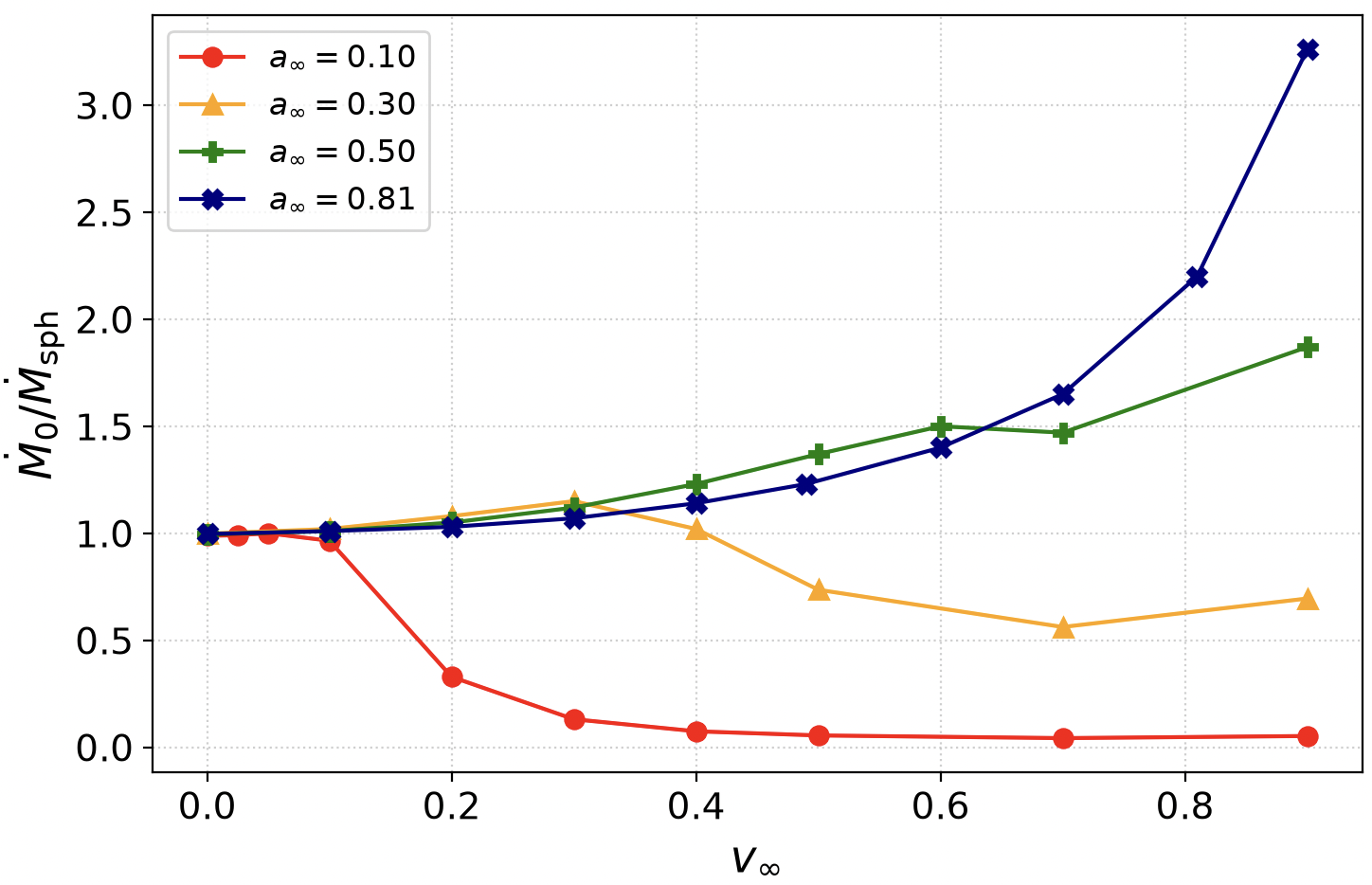}
        \includegraphics[width=0.48 \textwidth]{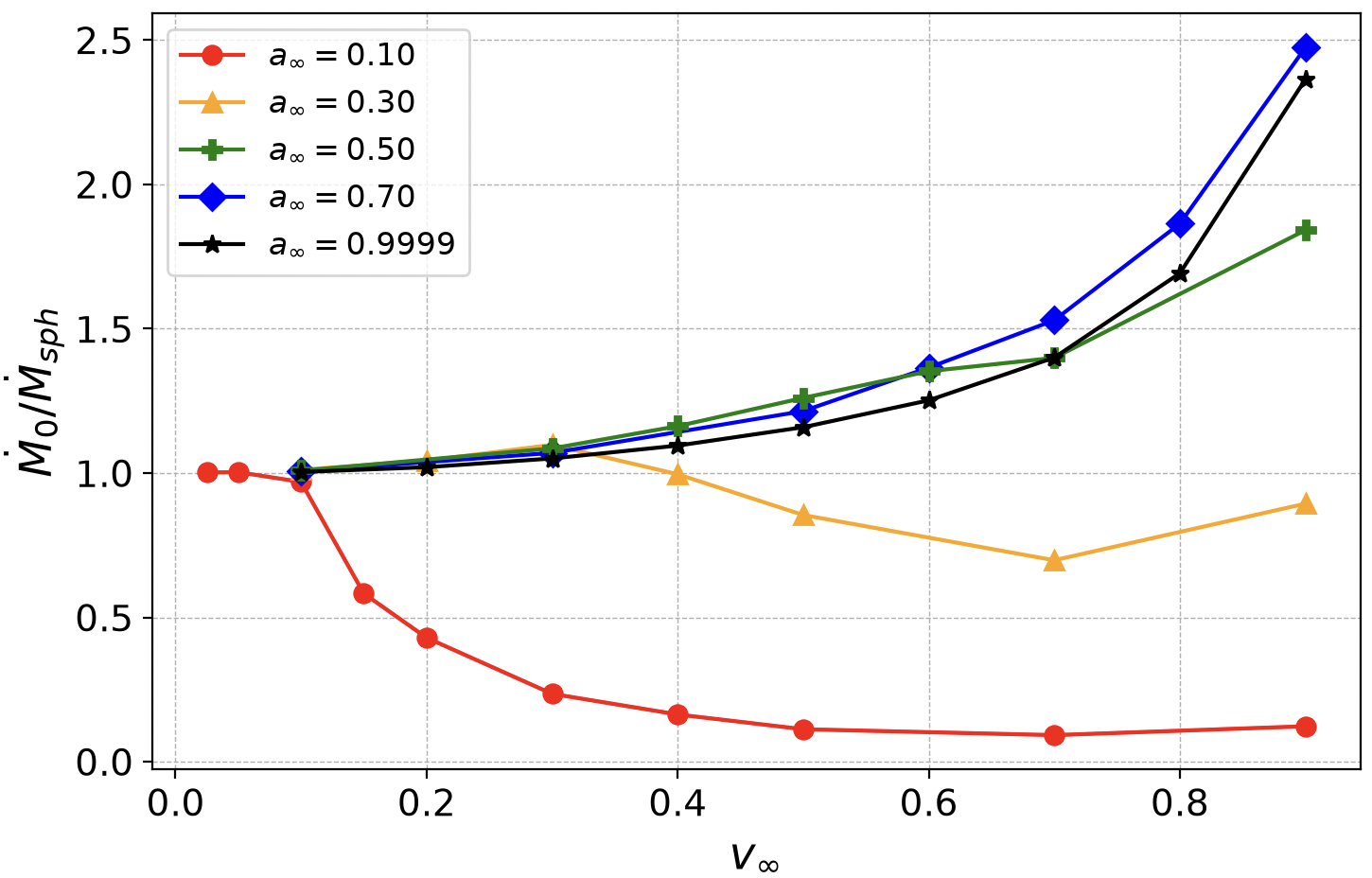}
        \caption{Rest-mass accretion rates $\dot M_0$, in units of $\dot M_{\rm sph}$, as a function of the relative speed $v_\infty$ for different values of the asymptotic sound speed $a_\infty$.  We show results for $\Gamma = 5/3$ in the left panel and for $\Gamma = 2$ in the right panel. The markers represent the numerical results listed in Table \ref{tab:GammaFiveThirds} and \ref{tab:GammaTwo}; the lines connecting the markers do not carry physical information, and are included as visual aids only.}
        \label{fig:seq}
\end{figure*}


        \label{fig:g2seq}

\subsection{Accretion rates}
\label{sec:res_accretion}

\begin{figure}[tb]
    \centering
    \includegraphics[width= 0.48\textwidth]{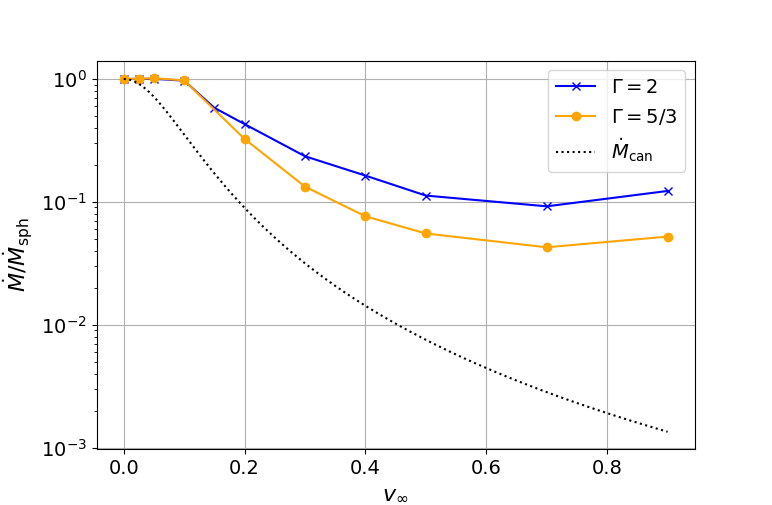}
    \caption{A comparison between our numerical results and the canonical accretion rate (\ref{eq:m_dot_can}) for $a_\infty = 0.1$.}
    \label{fig:m_can_comp}
\end{figure}

In Fig.~\ref{fig:seq} 
we graph our numerical results for the rest-mass accretion rate $\dot M_0$.  As we will discuss in more detail below, these accretion rates show not only quantitative but even qualitative differences from the canonical accretion rate (\ref{eq:m_dot_can}).   In order to highlight these differences we show in Fig.~\ref{fig:m_can_comp} a direct comparison for $a_\infty = 0.1$.

We first note that the canonical accretion rate (\ref{eq:m_dot_can}) predicts a monotonic decrease of $\dot M_0$ with increasing $v_\infty$.  Instead, we find that the accretion rates $\dot M_0$ {\em increase} with $v_\infty$ in the subsonic regime and again for large supersonic values of $v_\infty$.  The accretion rates decrease (with increasing $v_\infty$) only for mildly supersonic flow, and only for sufficiently small values of $a_\infty$. We observe the same qualitative behavior for both values of $\Gamma$. 

The origins of the discrepancy between the canonical accretion rate $\dot M_{\rm can}$ and the numerical results can be understood from the construction of $\dot M_{\rm can}$.  As we discussed in Sec.~\ref{sec:intro}, $M_{\rm can}$ serves as a crude estimate satisfying two criteria: (i) it approaches the spherical Bondi accretion rate (\ref{eq:m_dot_Bondi}) as $v_\infty = 0$, and (ii) it is similar to the Hoyle-Lyttleton accretion rate (\ref{eq:m_dot_HL}) for dust, i.e.~for $a_\infty = 0$.  Since the latter applies to supersonic flow only, the canonical accretion rate cannot be expected to accurately describe subsonic flow.  Moreover, since Hoyle-Lyttleton accretion is based on a purely Newtonian treatment, the canonical accretion rate also cannot account for relativistic effects.

%
\subsection{Shock and flow morphology}
\label{sec:num_shock}
%

 \begin{figure*}[t]
    \centering
    \includegraphics[width=1.0\textwidth]{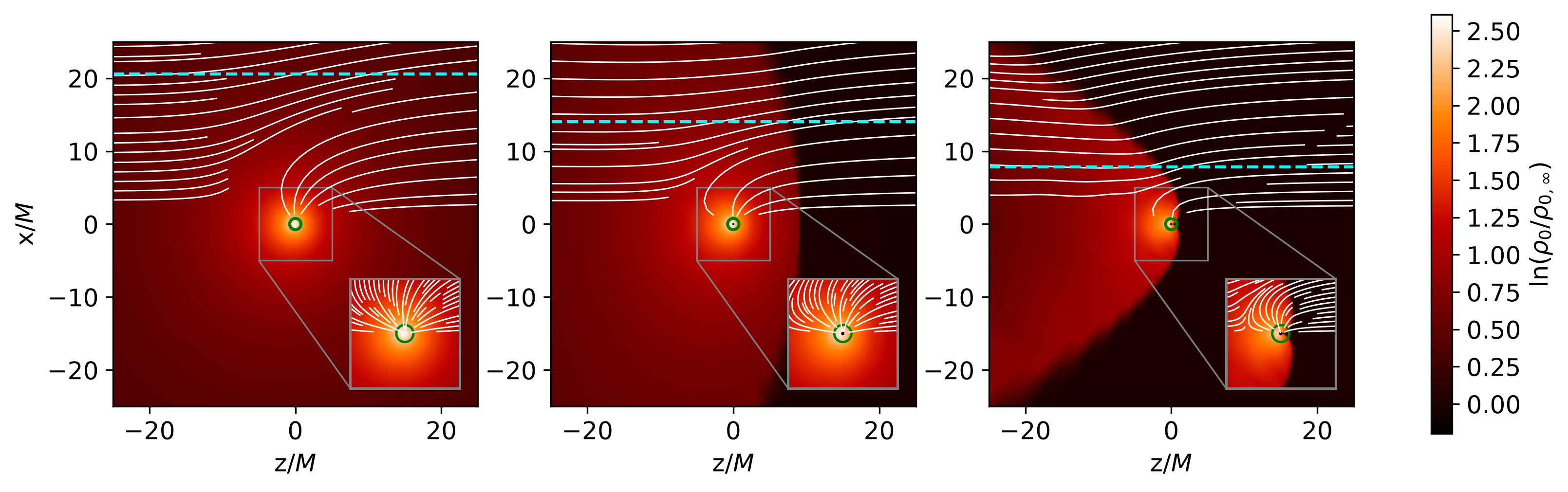}
    \caption{Rest mass density contours and streamline plots for different flow morphologies, for $\Gamma = 2$ and $a_\infty = 0.3$.  The left panel shows a subsonic example ($v_\infty = 0.2$), without the appearance of any shocks, the middle panel a mildly supersonic example ($v_\infty = 0.4$) that leads to the formation of a bow shock, and the right panel a more highly supersonic example ($v_\infty = 0.7$) that results in a shock cone (see Table \ref{tab:GammaTwo} for details.)  The green circles mark the location of the black hole horizons.  The white lines trace out the trajectories of fluid particles, and the dashed, horizontal cyan line marks the critical impact parameter $\zeta_{\rm crit}$ as computed from (\ref{eq:zeta_crit}).}
    \label{fig:ShockTypeExamples}
\end{figure*}

\begin{figure*}[t]
    \centering
    \includegraphics[width=0.48\textwidth]{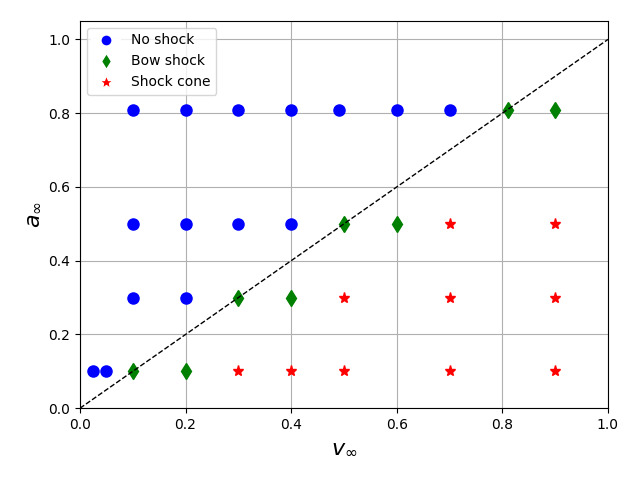}
    \includegraphics[width=0.48\textwidth]{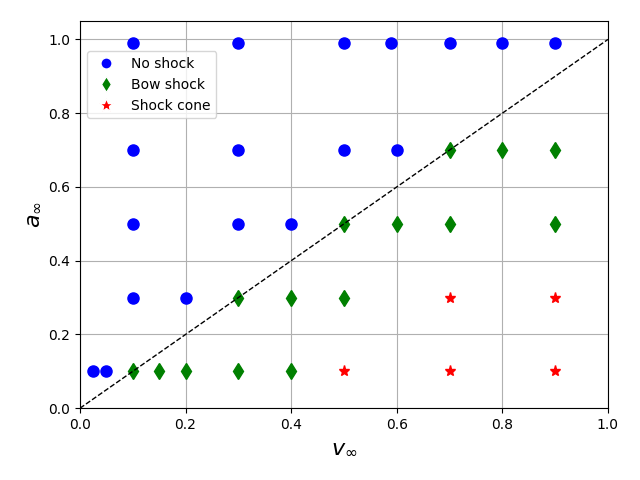}
    \caption{Shock types as a function of $v_\infty$ and $a_\infty$, for $\Gamma = 5/3$ in the left panel and $\Gamma = 2$ in the right panel.  The dashed lines mark $v_\infty = a_\infty$, i.e.~Mach number ${\mathcal M} = 1$.  For $\Gamma = 2$, the transition from a bow shock to a shock cone occurs at a larger value of ${\mathcal M}$ than for $\Gamma = 5/3$.}
    \label{fig:ShockType}
\end{figure*}



Several previous authors, starting with \cite{ShiMTS85}, have noted different types of shocks that may be encountered in the accretion onto a black hole moving through a fluid (see also the analytical treatment of \cite{FogR97}).   Here we relate the appearance of these different types of shocks to the non-monotonic dependence of the accretion rate on the relative speed $v_\infty$.

We show examples of the different types of shocks in Fig.~\ref{fig:ShockTypeExamples}, namely for $\Gamma = 2$ and $a_\infty = 0.3$.  In the left panel we show a subsonic example with $v_\infty = 0.2$, which does not develop any shocks.  The middle panel shows a mildly supersonic example for $v_\infty = 0.4$, which leads to a bow shock ahead of the black hole and detached from its horizon. Increasing $v_\infty$ results in the shock front approaching the black hole, until it becomes attached to the horizon.  We show an example of such a {\em shock cone}, for $v_\infty = 0.7$, in the right panel of Fig.~\ref{fig:ShockTypeExamples}. 

In Fig.~\ref{fig:ShockType} we mark the type of shock encountered for all combinations of $a_\infty$ and $v_\infty$ that we considered.  For both values of $\Gamma$, a bow shock first appears for $v_{\infty} = a_{\infty}$, marked by the dashed lines. For $\Gamma = 2$, i.e.~for the stiffer EOS, the bow shock remained detached from the horizon for larger values of $v_\infty$ than for $\Gamma = 5/3$.

Also shown in Fig.~\ref{fig:ShockTypeExamples} are flow lines for selected fluid particles, i.e.~integral lines of the fluid's four-velocity $u^a$.  In particular, we note that the flow lines appear to bend away from the symmetry axis when they cross a cone shock, which helps explain why the accretion rate decreases with the appearance of a cone shock.  In each panel in Fig.~\ref{fig:ShockTypeExamples} one can also identify a critical flow line that separates those fluid particles that get accreted by the black holes from those that do not.  Tracing back this critical fluid line to $z \rightarrow \infty$ would provide the critical impact parameter $\zeta_{\rm crit}$ first introduced below Eq.~(\ref{eq:m_dot_HL_flux}).  For comparison, we can independently compute $\zeta_{\rm crit}$ from the relativistic version of (\ref{eq:m_dot_HL_flux}), $\dot M_0 = \rho_{0\infty} v_\infty \gamma_\infty \pi \zeta_{\rm crit}^2$, i.e.
\begin{equation} \label{eq:zeta_crit}
\zeta_{\rm crit} = \left( \frac{\dot M_0}{\rho_{0\infty} v_\infty \gamma_\infty} \right)^{1/2},
\end{equation}
where $\dot M_0$ is computed numerically as discussed in Sect.~\ref{sec:num_restmass}.  In Fig.~\ref{fig:ShockTypeExamples} we include these values of $\zeta_{\rm crit}$ as the horizontal, dashed cyan  lines, and find good agreement with the critical flow lines.

%
\subsection{Simple analytical fits}
%

\begin{figure*}[t]
    \centering
    \includegraphics[width=0.48\textwidth]{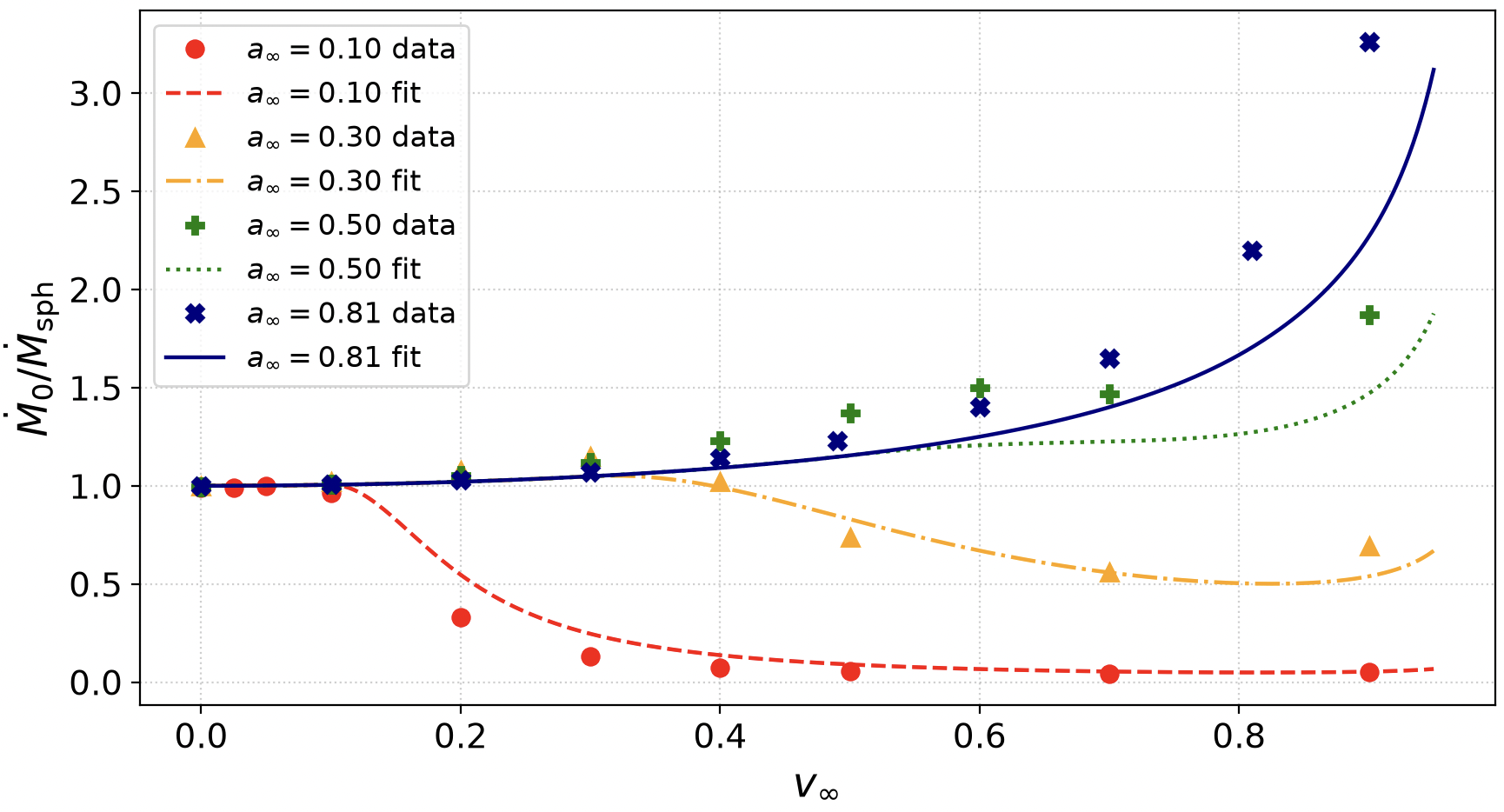}
    \includegraphics[width=0.48\textwidth]{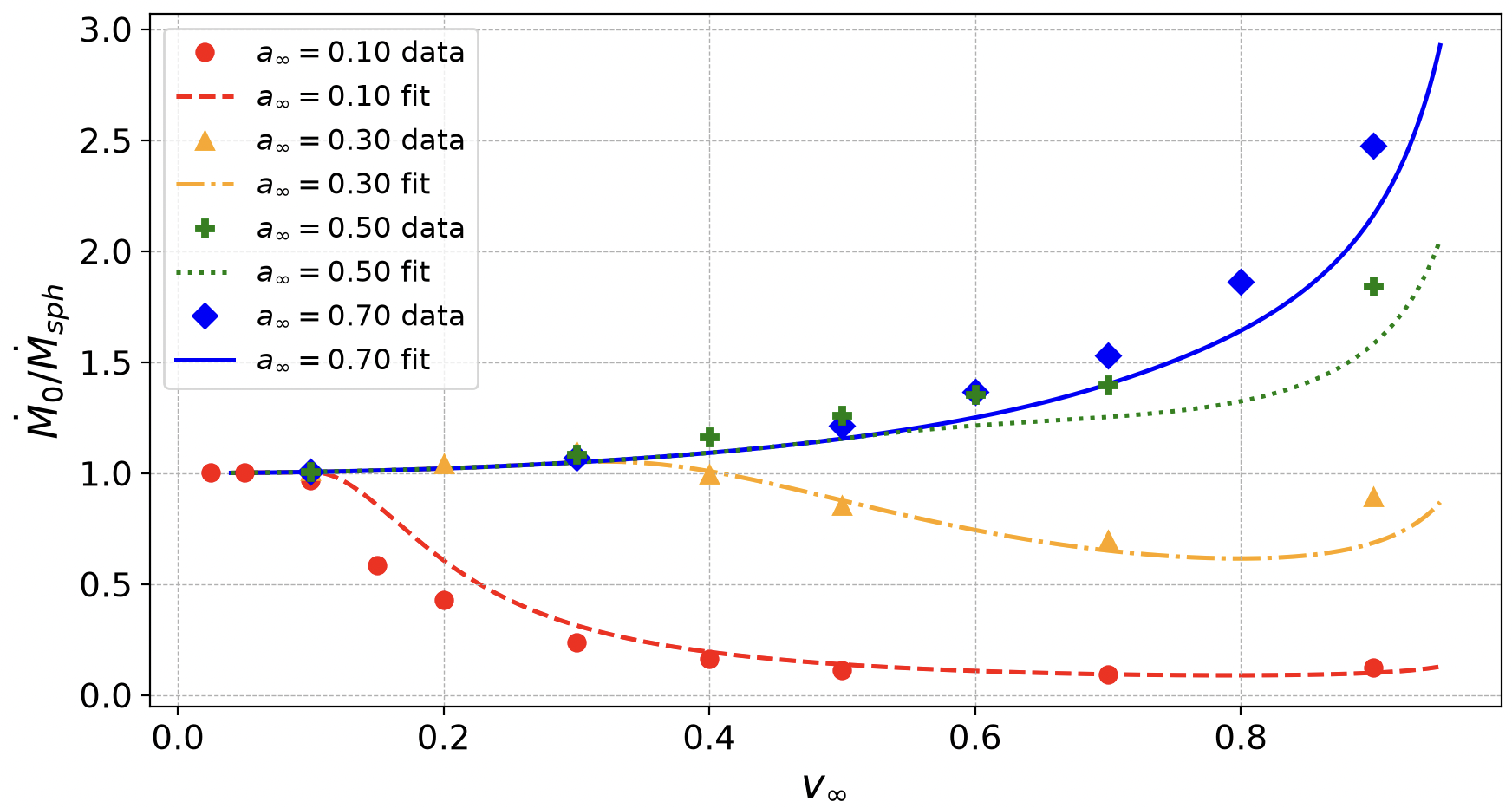}
    \caption{Comparison of numerical accretion rates (dots) with the simple analytical fit (\ref{eq:m_dot_fit}) (lines) for $\Gamma = 5/3$ with $q = 1.8$ (left panel) and for $\Gamma = 2$ with $q = 1.5$ (right panel).}
    \label{fig:fit}
\end{figure*}



In this Section we consider analytical fits that capture some of the qualitative features that we discussed above.  While more sophisticated fits have been considered by other authors (see, e.g., \cite{Ruf94}, albeit only in a Newtonian context), we here explore some simple generalizations of the canonical accretion rate (\ref{eq:m_dot_can}) that may provide analytical approximations suitable for applications like those discussed in Section \ref{sec:intro}.  

For subsonic flow, we observed that the accretion rate $\dot{M_0}$ deviates from the Bondi accretion rate $\dot M_{\rm sph}$ due to special relativistic effects only.  We therefore approximate $\dot M_0$ in this regime as
\begin{equation} \label{eq:M_dot_subsonic}
    \dot{M_0} = \gamma_{\infty} \dot{M}_{\rm sph} \quad (v_\infty \leq a_\infty).
\end{equation}
For subsonic Newtonian flow we have $\gamma_\infty = 1$, so that the accretion rate is independent of $v_\infty$ in this limit -- consistent with previous numerical simulations (see, e.g., \cite{Ruf94,PruGBPR24}).  For relativistic flow, the Lorentz factor $\gamma_\infty$ in (\ref{eq:M_dot_subsonic}) accounts for length contraction, and hence an enhance in the rest-mass density $\rho_0$ as measured by an observer moving with respect to the fluid.  The approximation (\ref{eq:M_dot_subsonic}) is also consistent with the extreme-relativistic analytical solution with $a_\infty = 1$ of \cite{PetST88}, which is always subsonic.  For that solution, the spherical accretion rate $\dot M_{\rm sph}$ can be found from (\ref{eq:m_dot_ur}) using $v_\infty = 0$ (and hence $\gamma_\infty = 1$), resulting in $\dot M_{\rm sph} = 16 \pi M^2 \rho_{0\infty}$ (see also Eq.~(57) in \cite{RicBS21a}).  Evidently,  (\ref{eq:M_dot_subsonic}) then yields the analytical result (\ref{eq:m_dot_ur}) exactly even for $v_\infty > 0$.

For supersonic flow, we model the drop in the accretion rate with a term inspired by the canonical accretion rate (\ref{eq:m_dot_can}), 
\begin{equation} \label{eq:M_dot_supersonic}
    \dot{M_0} = \gamma_{\infty}\dot{M}_{\rm sph}\left(\frac{a_{\infty}^2}{a_{\infty}^2 + (v_{\infty} - a_{\infty})^2}\right)^{q/2}
    \quad (v_\infty > a_\infty).
\end{equation}
Here the term $(v_{\infty} - a_{\infty})^2$ in the denominator ensures that $\dot M_0$ is continuous at the transition from subsonic to supersonic flow, but accounts for the drop in accretion rate observed for mildly supersonic flow.  In (\ref{eq:M_dot_supersonic}), $q$ is a dimensionless constant that is assumed to take the value $q=3$ in the canonical accretion rate (\ref{eq:m_dot_can}).  As discussed in \cite{BauS24c}, however, one might expect $q$ to be smaller for accretion of stiff fluids with $\Gamma > 5/3$ (see their Sec.~IV.B).  

We now combine the approximations (\ref{eq:M_dot_subsonic}) and (\ref{eq:M_dot_supersonic}) to obtain analytical fits of the form
\begin{equation} \label{eq:m_dot_fit}
  \dot{M} =
  \begin{cases}
    \gamma_{\infty}\,\dot{M}_{\rm{sph}}, 
      & v_{\infty} \le a_{\infty}, \\[6pt]
    \displaystyle
    \gamma_{\infty}\,\dot{M}_{\rm{sph}}
    \biggl(\frac{a_{\infty}^2}{a_{\infty}^2 + (v_{\infty} - a_{\infty})^2}\biggr)^{q/2},
      & v_{\infty} > a_{\infty}
  \end{cases}
\end{equation}
where $q$ is the only adjustable parameter.

In Fig.~\ref{fig:fit}  we show comparisons of our numerical values for accretion rates with the analytical fits (\ref{eq:m_dot_fit}) for $\Gamma = 5/3$ and $\Gamma = 2$ with $q = 1.8$ and $q = 1.5$, respectively.  While the agreement most certainly is not perfect, our fits (\ref{eq:m_dot_fit}), despite being very simple, do capture the qualitative features observed in the numerical results -- a slow and purely relativistic increase for subsonic flow, a marked drop for mildly supersonic flow, and an increase again as $v_\infty$ approaches the speed of light -- and generally reproduce the numerical values to within a factor of two or so.   

\subsection{Drag rates}
\label{sec:res_drag}

As discussed, for example, in \cite{PetSST89}, there are two main contributions to the drag rate exerted on the black hole moving through a fluid.  One contribution results from the momentum transferred to the black hole by the accreted material.  This contribution, referred to as the {\em capture} rate by \cite{PetSST89}, is given by
\begin{equation} \label{eq:drag_capture}
 \dot{p}^z_{\rm capt} = \frac{\rho_\infty + P_\infty}{\rho_{0\infty}} \dot M_0 \gamma_\infty v_\infty
\end{equation}
(see their Eq.~(2.40)).

Another contribution, referred to as the {\em deflection} rate by \cite{PetSST89}, results from long-range gravitational interactions.  These interactions result in a momentum transfer between the black hole and fluid particles that, while not accreted by the black hole, are deflected by its presence.  A number of different authors have estimated the resulting drag force in different regimes; we here adopt the Eq.~(B.45) of \cite{PetSST89},
\begin{equation} \label{eq:drag_deflect}
    \dot{p}^z_{\rm defl} = 4\pi(\rho_\infty + P_\infty)\frac{(1+2\gamma_\infty^2v_\infty^2)^2}{\gamma_\infty^2v_\infty^2}\ln\Lambda,
\end{equation}
where $\ln \Lambda=\ln(b_{\rm max}/b_{\rm min})$ is the Coulomb logarithm, given in terms of the fluid's minimum and maximum impact parameters $b_{\rm min}$ and $b_{\rm max}$.  Since neither one of these parameters can be determined without ambiguity, we treat $\ln \Lambda$ as a free parameter, roughly of order unity.

\begin{figure}[t]   
\centering
\includegraphics[width= 0.48\textwidth]{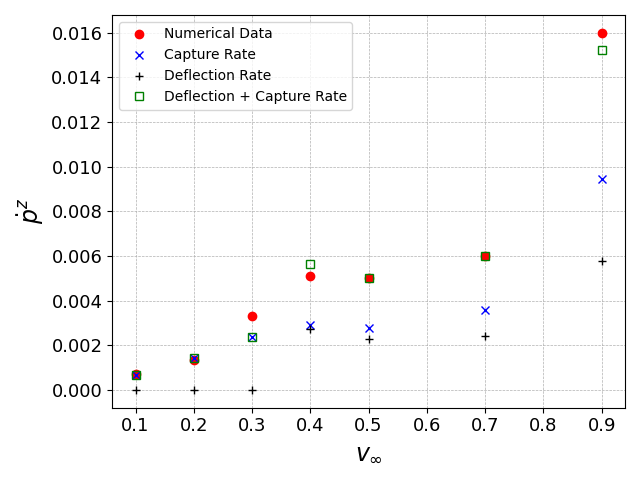}
    \caption{Drag rates $\dot p^z$ versus $v_\infty$ for $\Gamma = 5/3$ and $a_{\infty} = 0.3$. The red dots show our numerical results for $\dot{p^z}$, computed from the integral (\ref{eq:dragrate}).  The blue crosses show the capture rates (\ref{eq:drag_capture}), which we evaluate using the rest-mass accretion rate $\dot M_0$.  The black pluses show the estimated deflection drag rates (\ref{eq:drag_deflect}) -- which we assume to be non-zero for supersonic flow only -- using $\ln\Lambda = 1.2$.  The green stars, representing the sum of the capture and deflection rates, and hence our fits (\ref{eq:p_dot_fit}), agree well with numerical results.}
    \label{fig:capture_p}
\end{figure}

Several authors \cite{RepS80,PetSST89} have argued that the deflection drag can be ignored for subsonic flow, so that, for steady-state flow, the capture drag is the only contribution to the drag rate in this regime (but see also \cite{Ost99} for a more general analysis that takes into account the transition to steady-state).  As discussed in Appendix \ref{sec:Appendix_A_analytical}, the extreme-relativistic fluid accretion solution of \cite{PetSST89}, which is always subsonic, provides an analytical solution for which the drag rate (\ref{eq:dragrate}) indeed agrees exactly with the capture rate (\ref{eq:drag_capture}).  As another example we compare the different drag rates for a fluid with $\Gamma = 5/3$ and $a_\infty = 0.3$ in Fig.~\ref{fig:capture_p}.  For subsonic flow, our numerical results agree well with the capture rate (\ref{eq:drag_capture}) alone, while, for supersonic flow, the numerical results are reasonably well matched by a combination of the capture rate and the deflection rate (\ref{eq:drag_deflect}).

Motivated by these observations we now consider analytical fits for the drag rate that adopt the capture rate for subsonic flow only, and a sum of both the capture and deflection rates for supersonic flow,
\begin{widetext}
\begin{equation} \label{eq:p_dot_fit}
  \dot{p}^z =
  \begin{cases}
  \displaystyle 
\gamma_{\infty}v_\infty\frac{\rho_\infty+P_\infty}{\rho_{0\infty}}\dot M_0, 
      & v_{\infty} \le a_{\infty}, \\[6pt]
    \displaystyle
    \gamma_{\infty}v_\infty \frac{\rho_\infty+P_\infty}{\rho_{0\infty}} \dot M_0 +
    4\pi(\rho_\infty + P_\infty)\left(\frac{1+2\gamma_\infty^2v_\infty^2}{\gamma_\infty v_\infty} \right)^2\ln \Lambda & v_{\infty} > a_{\infty}.
  \end{cases}
\end{equation}
\end{widetext}
We note that these estimates are discontinuous at the transonic point $v_\infty = a_\infty$ (see also \cite{Ost99} for a discussion).  While we have not performed detailed simulations of the accretion in the immediate vicinity of the transonic point, our numerical data also suggest abrupt changes in the drag rate in its vicinity. 

\begin{figure*}[t]
    \centering
    \includegraphics[width=0.48\textwidth]{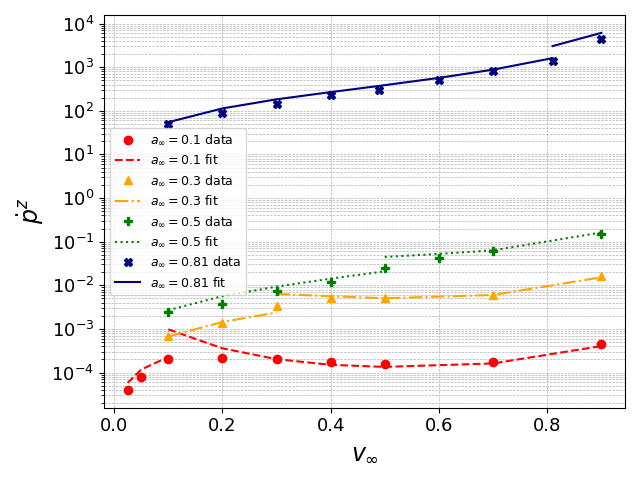}
    \includegraphics[width=0.48\textwidth]{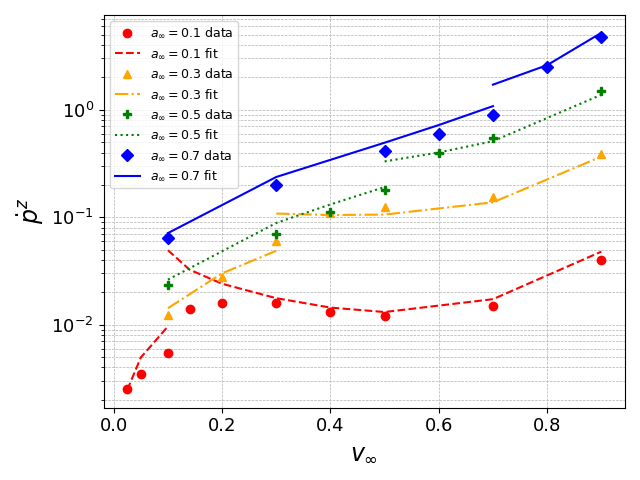}
    \caption{
    Drag rates $\dot p^z$ versus $v_\infty$ for $\Gamma = 5/3$ (left panel) and $\Gamma = 2$ (right panel). The icons show our numerical results while the lines show the estimated drag rates from (\ref{eq:p_dot_fit}) using $\ln\Lambda = 1.2$ for $\Gamma = 5/3$ and $\ln\Lambda = 0.6$ for $\Gamma = 2$.  Note that the estimated drag rates are discontinuous at the transonic point $v_\infty = a_\infty$. }
    \label{fig:P_plots}
\end{figure*}



In Fig.~\ref{fig:P_plots} we show numerical results for the drag rates $\dot p^z$ for $\Gamma = 5/3$ and $\Gamma = 2$ and compare with the fits (\ref{eq:p_dot_fit}).  Even though there are some noticeable differences between the numerical data and the fits, especially in the subsonic regime (see also the discussion in Appendix \ref{sec:Appendix_A_int}), overall the fits provide quite reasonable estimates for a large range of the parameter space, with a single choice of the free parameter $\ln \Lambda$ for each adiabatic exponent.

%
\section{Summary}
\label{sec:summary}
%

We reconsider relativistic accretion onto a Schwarzschild black hole that moves through an asymptotically uniform fluid.  Unlike many previous treatments (see, e.g., \cite{Edg04} for a review) we are especially interested in relativistic accretion of stiff fluids (with $\Gamma \geq 5/3$), which is particularly relevant for small black holes, potentially primordial in nature, that have been captured by neutron stars.  

We perform numerical simulations adopting the Cowling approximation, by which we assume that the black hole spacetime is unaffected by the fluid. In our simulation, we constructed this black hole spacetime in maximally sliced trumpet coordinates, which penetrate the black hole horizon and hence allow for a smooth fluid flow into the black hole.

We measure the accretion and drag rates for different values of the asymptotic sound speed $a_\infty$ and the asymptotic relative speed $v_\infty$ for both $\Gamma = 5/3$ and $\Gamma = 2$.  For both values of $\Gamma$ we observe similar characteristic features.  For subsonic flow, we observe an increase in the accretion rate that is consistent with the increase of the Lorentz factor $\gamma_{\infty} = (1 - v_{\infty}^2)^{-1/2}$. For (mildly) supersonic flow the appearance of shocks leads to a decrease in the accretion rate, while for highly relativistic flow, with the relative speed $v_\infty$ approaching the speed of light, the Lorentz factor again leads to an increase in the accretion rate.   The commonly adopted ``canonical" Bondi-Hoyle-Lyttleton approximation (\ref{eq:m_dot_can}) fails to capture these features and leads to large discrepancies between the numerical and predicted accretion rates, even for subsonic and Newtonian flow (see Fig.~\ref{fig:m_can_comp}).  We propose simple generalizations of the canonical accretion rate that (i) reproduce the above features qualitatively, and (ii) generally agree with the numerical values to within a factor of two or better (see Fig.~\ref{fig:fit}).  We similarly consider fits for the drag rates -- consisting of contributions from both the capture and deflection rates -- and find reasonable agreement across a large range of parameter space (see Fig.~\ref{fig:P_plots}).

\acknowledgments

EE and PH were supported by undergraduate research fellowships at Bowdoin College.  This work was supported in parts by National Science Foundation (NSF) grants PHY-2308821 and PHY-2605218 to Bowdoin College, as well as NSF grant PHY-2308242 to the University of Illinois at Urbana-Champaign.  

%
\appendix

\section{Drag Rates}
\label{sec:Appendix_A}
%
\subsection{Surface integration for drag rates}
\label{sec:Appendix_A_int}

In order to calculate the drag force on the black hole we evaluate the surface integral (\ref{eq:dragrate}).  Given a fixed, spherically symmetric background metric with spatial metric $\gamma_{ij} = \psi^4 \, \eta_{ij}$, where $\psi$ is the conformal factor and $\eta_{ij}$ the flat metric in spherical polar coordinates, the integral (\ref{eq:dragrate}), evaluated on a large sphere of radius $r_{\rm int}$ centered on the black hole, can be written as
\begin{align} \label{eq:drag_rate}
    \dot{p}^z & = - \oint_{r_{\rm int}} T^{zi}\sqrt{-g}dS_i = -{\int}_{r_{\rm int}}T^{zr}\sqrt{-g}d{\theta}d{\phi} \nonumber \\
    & = -2{\pi}{\alpha}r_{\rm int}^2 \psi^6{\int}_{r_{\rm int}}{\sin{\theta}}T^{zr}d{\theta},
\end{align}
where $\alpha$ is the lapse function.  We then calculate the mixed Cartesian-spherical component $T^{zr}$ of the stress-energy tensor 
\begin{equation}
    T^{\mu\nu} = (\rho + P)u^{\mu}u^{\nu} + Pg^{\mu\nu},
\end{equation}
using the transformation
\begin{equation}
    T^{zr} = \cos{\theta}T^{rr} - r\sin{\theta}T^{r\theta}
\end{equation}
and compute the components in spherical polar coordinates from the hydrodynamical variables used in the code.

Unlike the integral (\ref{eq:accretionrate}) for the rest-mass accretion rate $\dot M_0$, which can be evaluated over any closed surface once steady-state has been achieved, the integral (\ref{eq:dragrate}) for the drag rate is accurate only in the limit $r_{\rm int} \rightarrow \infty$ (see the discussion in Sec.~\ref{sec:res_drag}).  We would therefore like to evaluate the drag rate at as late a time $t_{\rm int}$ as possible (so that the numerical evolution has settled down to steady-state) and at a radius $r_{\rm int}$ as large as possible (so that the integral (\ref{eq:dragrate}) becomes accurate).  For subsonic flow, however, there is the additional complication that numerical artifacts from the (down-stream) outer boundary can propagate into the numerical domain.  Therefore, we evaluate the drag rate at the latest time available and at $r_{\rm int} = r_{\rm max}$ for supersonic flow, but restrict values of $t_{\rm int}$ and $r_{\rm int}$ to combinations that are not yet affected by our approximate outer boundaries for subsonic flow.  Because of these limitations, our numerical results for drag rates are more affected by numerical error than the rest-mass accretion rates, especially for subsonic flow.

\subsection{An analytical solution}
\label{sec:Appendix_A_analytical}

The authors of \cite{PetST88} provide an analytical solution for the accretion of an extreme-relativistic fluid with $P = \rho$, hence $\Gamma = 2$ and $a_\infty = 1$, onto a moving and possibly spinning black hole.  For accretion onto a Schwarzschild black hole (with zero spin), the flow profile is given by their Eqs.~(19) and (20).   Inserting these profiles, evaluated in the limit $r \rightarrow \infty$, into equation (\ref{eq:drag_rate}) and integrating we find 
\begin{equation} \label{eq:p_dot_ur}
    \dot{p}^z = 32\pi{M^2} \gamma_\infty^2 {v_\infty}\rho_\infty.
\end{equation}
This analytical solution now provides a valuable test for our numerical evaluation of drag rates, as shown in Fig.~\ref{fig:ur_drag}. 

We note that Eq.~(\ref{eq:p_dot_ur}) matches the expression of \cite{PetSST89} for the capture contribution to drag force identically.  Specifically, inserting (\ref{eq:m_dot_ur}) for $\dot M_0$ in (\ref{eq:drag_capture}) and using $\rho_\infty = P_\infty$ for the extreme-relativistic fluid results in (\ref{eq:p_dot_ur}).  Accretion of an extreme-relativistic fluid, which is always subsonic, therefore provides an example in which the drag force given by entirely by the capture rate, without any contributions from the deflection rate (compare the discussion in Sec.~\ref{sec:res_drag}).


\end{document}